\documentclass[12pt]{article}

\usepackage{epsfig}
\usepackage{graphicx}
\usepackage{dcolumn}
\usepackage{bm}

\date{}
\begin{document}
%\documentstyle[12pt]{article}

%\documentclass[preprint]{revtex4}

% Some other (several out of many) possibilities
%\documentclass[preprint,aps]{revtex4}
%\documentclass[preprint,aps,draft]{revtex4}
%\documentclass[prc]{revtex4}% Physical Review C

%\begin{document}
\newcommand{\mc}{\multicolumn}
\newcommand{\bce}{\begin{center}}
\newcommand{\ece}{\end{center}}
\newcommand{\beq}{\begin{equation}}
\newcommand{\eeq}{\end{equation}}
\newcommand{\bea}{\begin{eqnarray}}

\newcommand{\eea}{\end{eqnarray}}
\newcommand{\cont}{\nonumber\eea\bea}
\newcommand{\cl}[1]{\begin{center} {#1} \end{center}}
\newcommand{\ba}{\begin{array}}
\newcommand{\ea}{\end{array}}
%\newcommand{\arr}{\bea}

% -------------- MATH def -------------------------------
\newcommand{\ab}{{\alpha\beta}}
\newcommand{\cd}{{\gamma\delta}}
\newcommand{\dc}{{\delta\gamma}}
\newcommand{\ac}{{\alpha\gamma}}
\newcommand{\bd}{{\beta\delta}}
\newcommand{\abc}{{\alpha\beta\gamma}}
\newcommand{\eps}{{\epsilon}}
\newcommand{\lam}{{\lambda}}
\newcommand{\mn}{{\mu\nu}}
\newcommand{\mpnp}{{\mu'\nu'}}
\newcommand{\Amuu}{{A_{\mu}}}
\newcommand{\Amuo}{{A^{\mu}}}
\newcommand{\Vmuu}{{V_{\mu}}}
\newcommand{\Vmuo}{{V^{\mu}}}
\newcommand{\Anuu}{{A_{\nu}}}
\newcommand{\Anuo}{{A^{\nu}}}
\newcommand{\Vnuu}{{V_{\nu}}}
\newcommand{\Vnuo}{{V^{\nu}}}
\newcommand{\Fmnu}{{F_{\mu\nu}}}
\newcommand{\Fmno}{{F^{\mu\nu}}}

\newcommand{\abcd}{{\alpha\beta\gamma\delta}}

% Boldmath definitions

\newcommand{\bsigma}{\mbox{\boldmath $\sigma$}}
\newcommand{\btau}{\mbox{\boldmath $\tau$}}
\newcommand{\brho}{\mbox{\boldmath $\rho$}}
\newcommand{\bpipi}{\mbox{\boldmath $\pi\pi$}}
\newcommand{\bss}{\bsigma\!\cdot\!\bsigma}
\newcommand{\btt}{\btau\!\cdot\!\btau}
\newcommand{\bnabla}{\mbox{\boldmath $\nabla$}}
\newcommand{\bphi}{\mbox{\boldmath $\tau$}}
\newcommand{\bvarphi}{\mbox{\boldmath $\rho$}}
\newcommand{\bDelta}{\mbox{\boldmath $\Delta$}}
\newcommand{\bpsi}{\mbox{\boldmath $\psi$}}
\newcommand{\bPsi}{\mbox{\boldmath $\Psi$}}
\newcommand{\bPhi}{\mbox{\boldmath $\Phi$}}
\newcommand{\bnab}{\mbox{\boldmath $\nabla$}}
\newcommand{\bpi}{\mbox{\boldmath $\pi$}}
\newcommand{\btheta}{\mbox{\boldmath $\theta$}}
\newcommand{\bkappa}{\mbox{\boldmath $\kappa$}}

\newcommand{\bA}{{\bf A}}
\newcommand{\bfe}{{\bf e}}
\newcommand{\bb}{{\bf b}}
\newcommand{\br}{{\bf r}}
\newcommand{\bj}{{\bf j}}
\newcommand{\bk}{{\bf k}}
\newcommand{\bl}{{\bf l}}
\newcommand{\bL}{{\bf L}}
\newcommand{\bM}{{\bf M}}
\newcommand{\bp}{{\bf p}}
\newcommand{\bq}{{\bf q}}
\newcommand{\bR}{{\bf R}}
\newcommand{\bs}{{\bf s}}
\newcommand{\bS}{{\bf S}}
\newcommand{\bT}{{\bf T}}
\newcommand{\bv}{{\bf v}}
\newcommand{\bV}{{\bf V}}
\newcommand{\bx}{{\bf x}}
\newcommand{\fph}{${\cal F}$}
\newcommand{\aph}{${\cal A}$}
\newcommand{\dph}{${\cal D}$}
\newcommand{\fpi}{f_\pi}
\newcommand{\mpi}{m_\pi}
\newcommand{\Tr}{{\mbox{\rm Tr}}}
\def\Qb{\overline{Q}}
\newcommand{\delu}{\partial_{\mu}}
\newcommand{\delo}{\partial^{\mu}}
%\newcommand{\half}{{1\over 2}}
%\newcommand{\quart}{{1\over 4}}
%
%
% ------------------ arrow mod ---------------------
\newcommand{\up}{\!\uparrow}
\newcommand{\upup}{\uparrow\uparrow}
\newcommand{\updo}{\uparrow\downarrow}
\newcommand{\uu}{$\uparrow\uparrow$}
\newcommand{\ud}{$\uparrow\downarrow$}
\newcommand{\auu}{$a^{\uparrow\uparrow}$}
\newcommand{\aud}{$a^{\uparrow\downarrow}$}
\newcommand{\pu}{p\!\uparrow}

% ------------------------------------------------------
\newcommand{\qp}{quasiparticle}
\newcommand{\sa}{scattering amplitude}
\newcommand{\ph}{particle-hole}
\newcommand{\qcd}{{\it QCD}}
\newcommand{\integ}{\int\!d}
\newcommand{\ie}{{\sl i.e.~}}
\newcommand{\etal}{{\sl et al.~}}
\newcommand{\etc}{{\sl etc.~}}
\newcommand{\rhs}{{\sl rhs~}}
\newcommand{\lhs}{{\sl lhs~}}
\newcommand{\eg}{{\sl e.g.~}}
\newcommand{\ef}{\epsilon_F}
\newcommand{\sigt}{d^2\sigma/d\Omega dE}
\newcommand{\sige}{{d^2\sigma\over d\Omega dE}}
% ----------------------- ------------------------------
\newcommand{\rpaeq}{\beq
\left ( \begin{array}{cc}
A&B\\
-B^*&-A^*\end{array}\right )
\left ( \begin{array}{c}
X^{(\kappa})\\Y^{(\kappa)}\end{array}\right )=E_\kappa
\left ( \begin{array}{c}
X^{(\kappa})\\Y^{(\kappa)}\end{array}\right )
\eeq}
\newcommand{\ket}[1]{| {#1} \rangle}
\newcommand{\bra}[1]{\langle {#1} |}
\newcommand{\ave}[1]{\langle {#1} \rangle}

%\newcounter{f1}
%\newcounter{f2}
%\renewcommand{\theequation}{\thesubsection.\arabic{equation}}
%\renewcommand{\thetable}{\thesection.\arabic{table}}
\newcommand{\singlespace}{
    \renewcommand{\baselinestretch}{1}\large\normalsize}
\newcommand{\doublespace}{
    \renewcommand{\baselinestretch}{1.6}\large\normalsize}
\newcommand{\bftau}{\mbox{\boldmath $\tau$}}
\newcommand{\bfalpha}{\mbox{\boldmath $\alpha$}}
\newcommand{\bfgamma}{\mbox{\boldmath $\gamma$}}
\newcommand{\bfxi}{\mbox{\boldmath $\xi$}}
\newcommand{\bfbeta}{\mbox{\boldmath $\beta$}}
\newcommand{\bfeta}{\mbox{\boldmath $\eta$}}
\newcommand{\bfpi}{\mbox{\boldmath $\pi$}}
\newcommand{\bfphi}{\mbox{\boldmath $\phi$}}
\newcommand{\bfR}{\mbox{\boldmath ${\cal R}$}}
\newcommand{\bfL}{\mbox{\boldmath ${\cal L}$}}
\newcommand{\bfM}{\mbox{\boldmath ${\cal M}$}}
\def\dblint{\mathop{\rlap{\hbox{$\displaystyle\!\int\!\!\!\!\!\int$}}
    \hbox{$\bigcirc$}}}
\def\ut#1{$\underline{\smash{\vphantom{y}\hbox{#1}}}$}

\def\UNITY{{\bf 1\! |}}
\def\Pom{{\bf I\!P}}
\def\lsim{\mathrel{\rlap{\lower4pt\hbox{\hskip1pt$\sim$}}
    \raise1pt\hbox{$<$}}}         %less than or approx. symbol
\def\gsim{\mathrel{\rlap{\lower4pt\hbox{\hskip1pt$\sim$}}
    \raise1pt\hbox{$>$}}}         %greater than or approx. symbol
\def\beq{\begin{equation}}
\def\eeq{\end{equation}}
\def\bea{\begin{eqnarray}}
\def\eea{\end{eqnarray}}

%\doublespace

\begin{center}  
{\Large\bf  Lectures on Diffraction and Saturation 
         of \vspace{0.5cm}\\

Nuclear Partons in DIS off Heavy Nuclei
\footnote{The notes of lectures presented by N.N.N. at the
XXXVI St.Petersburg Nuclear Physics Institute
Winter
    School on Nuclear and Particle Physics \& VIII St.Petersburg
    School on Theoretical Physics,
St.Petersburg, Repino, February 25 - March 3, 2002} }
\\ \vspace{1cm}
 { \bf I.P. Ivanov$^{a}$, {\underline{  N.N. Nikolaev}}$^{b,c)}$,
W. Sch\"afer$^{b)}$, B.G. Zakharov$^{c)}$ \medskip\\

\& V.R.
Zoller$^{d)}$\bigskip\\  }
{\small \it
$^{a)}$ Institute of Mathematics, Novosibirsk, Russia\\
$^{b)}$ Institut f. Kernphysik, Forschungszentrum J\"ulich, D-52425 J\"ulich, Germany\\
$^{c)}$ L.D.Landau Institute for Theoretical Physics, Chernogolovka, Russia\\
$^{d)}$ Institute for Theoretical and Experimental Physics, Moscow, Russia\\
E-mail: N.Nikolaev$@$fz-juelich.de\vspace{1cm} \\}

% $\UNITY$
{\bf Abstract\\    }

\end{center}
{\small The Lorentz contraction of ultrarelativistic nuclei entails
a spatial overlap and fusion (recombination, saturation) of 
partons belonging to different nucleons at the same impact 
parameter. In these lectures we present a consistent description 
of the fusion of partons in terms of nuclear attenuation of 
color dipole states of the photon and collective
Weizs\"acker-Williams (WW) gluon structure function of a nucleus.
The point that all observables for DIS off nuclei are 
uniquely calculable in terms of the nuclear WW glue amounts to
a new form of factorization in the saturation regime. \\
 We start with the theory 
of multichannel propagation of color dipoles in a nuclear medium including
the color-singlet to color-octet to color-octet transitions. 
We show how the Glauber-Gribov formulas are recovered from 
the multichannel formalism. Then we  derive the two-plateau 
momentum distribution of final state (FS) quarks produced in  
deep inelastic scattering (DIS) off
nuclei in the saturation regime. The diffractive plateau which
dominates for small $\bp$ measures precisely the momentum
distribution of quarks in the beam photon, the r\^ole of the
nucleus is simply to provide an opacity. The plateau for truly
inelastic DIS exhibits a substantial nuclear broadening of the FS
momentum distribution.  The Weizs\"acker-Williams glue of a
nucleus exhibits a substantial nuclear dilution, still soft 
initial state (IS)
nuclear sea saturates because of the anti-collinear splitting of
gluons into sea quarks. Then we comment on the signatures of
saturation in exclusive diffractive DIS.
\\
A large body of 
these lectures is on the recent theory of 
jet-jet inclusive cross sections. We 
show that for hard
dijets the decorrelation momentum is of the order
of the nuclear saturation momentum $Q_A$. For minijets with the
transverse momentum below the saturation scale we predict a
complete disappearance of the azimuthal jet-jet correlation. We conclude
with comment on a possible 
relevance of the decorrelation of jets to the experimental data 
from the STAR-RHIC Collaboration. \medskip\\}

%\doublespace

% ------------------------  Section 1

\section{Introduction}

Within the standard QCD parton model the virtual photoabsorption
cross section is proportional to the density of partons in the
target. Inverting the argument, one would define the density 
of partons in a target $A$ as
\beq
F_2(x,Q^2) = \sum_{f} e_f^2 [q_f(x,Q^2) + \bar{q}_f(x,Q^2)] =
{Q^2 \over 4\pi^2 \alpha_{em}} \sigma_{\gamma^*A}
\label{eq:1.1}
\eeq
In the so-called brick-wall or Breit frame, in which 
the photon has a zero energy and the (target) hadron or nucleus 
is ultrarelativistic, DIS amounts to backward reflection of the
parton with the longitudinal momentum $k_z = xP_{z}= {1\over 2}q_z= 
{1\over 2}\sqrt{Q^2}$,
see fig.1. 

% -----------  figure 1

\begin{figure}[!htb]
   \centering
   \epsfig{file=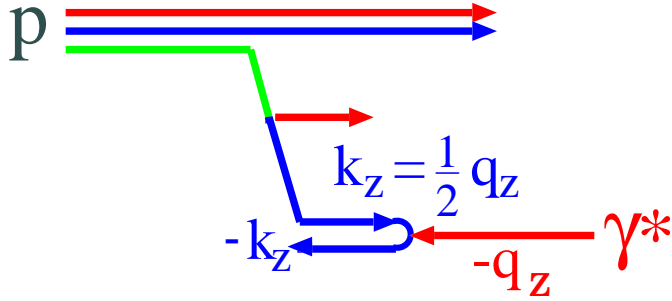,width=9cm}
\caption{\it DIS in the Breit frame}
\end{figure}
\noindent

When DIS is viewed in the laboratory frame, the 
hadronic properties of photons suggest \cite{NZfusion} a nuclear shadowing by
which the density of partons in a nucleus rises slower than
$\propto A$. Very heavy nuclei will be opaque to a part of
the Fock states of the photon \cite{NZZdiffr}. 
When viewed in the Breit frame,
the Lorentz contraction of an ultrarelativistic  
nucleus entails a spatial 
overlap and fusion of partons at
\beq
x \lsim x_A={1\over R_A m_N}
\label{eq:1.2}
\eeq
as indicated in fig.~2.

% -----------  figure 2
\begin{figure}[!htb]
   \centering
   \epsfig{file=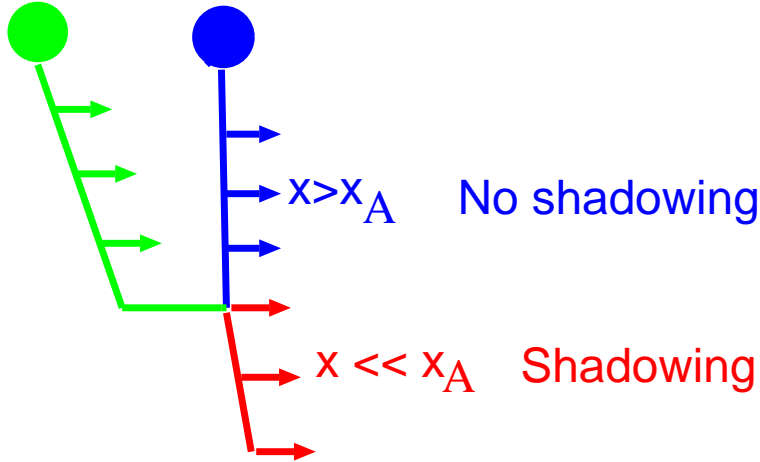,width=10cm}
\caption{\it The fusion of partons from overlapping parton clouds 
of two nucleons in a Lorentz-contracted nucleus}
\end{figure}
This interpretation of nuclear opacity in terms of a fusion
and saturation of nuclear partons has been introduced in
1975 \cite{NZfusion} way before the QCD parton model. 
The review of early works on nuclear shadowing
can be found in lectures by one of the authors 
at the 1976 Winter School of PNPI
\cite{LIYAF76}, see also \cite{UFN}.
The pQCD link between nuclear opacity and
saturation has been considered in ref. \cite{NZ91} and 
by Mueller \cite{Mueller1},  the
pQCD discussion of fusion of nuclear gluons has been revived by
McLerran et al. \cite{McLerran}.

The common wisdom is that in DIS the final state (FS) interaction 
of the struck parton, which in the Breit frame moves in the 
direction opposite to the debris of the nucleon/nucleus, see fig.~2, 
can be
neglected and the observed transverse and longitudinal 
momentum distribution of struck partons
in the FS coincides with the initial state (IS) 
density of partons in the probed
hadron. Based on the consistent treatment of intranuclear
distortions, we derive the two-plateau spectrum of FS quarks
found in \cite{Saturation}. Following our
early analysis of diffractive DIS off nucleons \cite{NZ92,NZsplit}
and nuclei \cite{NSSdijet}, we introduce the collective 
Weizs\"acker-Williams gluon structure function of the nucleus,
which uniquely defines all nuclear DIS observables. 
We find a substantial nuclear broadening of inclusive FS spectra and
demonstrate that despite this broadening the FS sea parton density
exactly equals the IS sea parton density calculated in terms of
the WW glue of the nucleus as defined according to \cite{NSSdijet}. We
pay special attention to an important point that diffractive DIS
in which the target nucleus does not break and is retained in the
ground state, makes precisely 50 per cent of the total DIS events
for heavy nuclei at small $x$ \cite{NZZdiffr}. 
We point out that the saturated diffractive
plateau measures precisely the momentum distribution of
(anti-)quarks in the $q\bar{q}$ Fock state of the photon. In
contrast to DIS off nuclei, the fraction of DIS off free nucleons
which is diffractive, and which also measures unitarity corrections
to the two-gluon exchange approximation \cite{NZ94,BGNPZunit},
 is negligibly small \cite{GNZ95}, $\eta_D \lsim
$ 6-10 \%, and there is little room for genuine saturation effects
even at HERA. We show how the anti-collinear splitting of WW
gluons into sea quarks gives rise to nuclear saturation of the sea
despite a substantial nuclear dilution of the WW glue.

After establishing the properties of the single-jet inclusive cross
section, we turn our attention to the recent theory of 
jet-jet (de)correlations \cite{JetJet}.
After a hard scattering partons with large transverse momentum
fragment into a high-energy jet of particles. In deep inelastic
scattering at small $x$, to the collinear approximation, the 
photon-gluon fusion $\gamma^* g \to q\bar{q}$, often referred 
to as an interaction of the unresolved 
photon, gives rise to the back-to-back jets in the current
or photon fragmentation region. The experimental signature of
the unresolved photon
interaction is that the so-called lightcone plus-components of the jet
momenta sum up to the lightcone plus-component of the photon's 
momentum. The allowance for the 
transverse momentum of gluons leads to the disparity of the 
momenta and azimuthal decorrelation of the quark and antiquark
jets, which within the $k_{\perp}$-factorization can be quantified 
in terms of the so-called unintegrated gluon structure function of 
the target (see \cite{Azimuth} and references therein).

In view of the substantial nuclear broadening
of the unintegrated gluon SF of nuclei it is natural to
expect stronger azimuthal decorrelation of jets produced in DIS 
off nuclei. In these lectures we report a consistent
multiple-scattering 
theory of jet-jet decorrelation in the genuinely 
inelastic DIS followed by color excitation 
of the target nucleus. For heavy nuclei of equal importance is  
coherent diffractive DIS in which the target nucleus does not break 
and is retained in the ground state. In 
these coherent diffractive events quark and antiquark jets are
produced exactly back-to-back with negligibly small transverse 
decorrelation momentum 
\beq
|\bDelta|=
|\bp_+ + \bp_-| \lsim 1/R_A \sim m_{\pi}/A^{1/3}\,,
\label{eq:1.3}
\eeq 
where $R_A$ is the radius of the target nucleus. 
For hard jets diffractive attenuation 
effects are weak and we report a compact formula for the broadening 
of azimuthal correlations between the quark and antiquark jets.
In this case the decorrelation (acoplanarity, out-of-plane) momentum is given
by the nuclear saturation momentum $Q_A$. Then we prove
a disappearance of the jet-jet correlation for minijets
with the transverse momentum below the saturation scale $Q_A$. 
In the Conclusions we summarize our principal findings and 
speculate on their relevance to the recent observation
of the dissolution of the away jets in central nuclear collisions 
at RHIC \cite{STARRHIC}. 

% -----------------  Section 2

\section{Quark and antiquark jets in 
DIS off free nucleons: single particle spectrum and jet-jet decorrelation}

We recall briefly the color dipole formulation of DIS
\cite{NZZdiffr,NZ91,NZ92,NZglue,NZ94,NZZlett} and illustrate 
our ideas on an example of jet-jet decorrelation in DIS
off free nucleons which at moderately small $x$ is dominated by
interactions of $q\bar{q}$ states of the photon. We evaluate
the jet production cross section at the parton level. The total cross
section for interaction of the color dipole $\br$ with the target
nucleon equals 
\bea 
\sigma(r)&=& \alpha_S(r) \sigma_0\int d^2\bkappa
f(\bkappa )\left[1 -\exp(i\bkappa \br )\right]\nonumber\\
& = &
{1\over 2}\alpha_S(r) \sigma_0\int d^2\bkappa
f(\bkappa )\left[1 -\exp(i\bkappa \br )\right]
\cdot \left[1 -\exp(-i\bkappa \br )\right]
\, , 
\label{eq:2.1}
\eea 
where $f(\bkappa )$ is related to the unintegrated glue of
the target nucleon by 
\bea 
f(\bkappa ) = {4\pi \over
N_c\sigma_0}\cdot {1\over \kappa^4} \cdot {\partial G \over
\partial\log\kappa^2}
\label{eq:2.2}
\eea
and is normalized as
\bea
\int d^2\bkappa  f(\bkappa )=1\, .
\label{eq:2.3} 
\eea

% -----------  figure 3

\begin{figure}[!htb]
\begin{center}
\epsfig{file=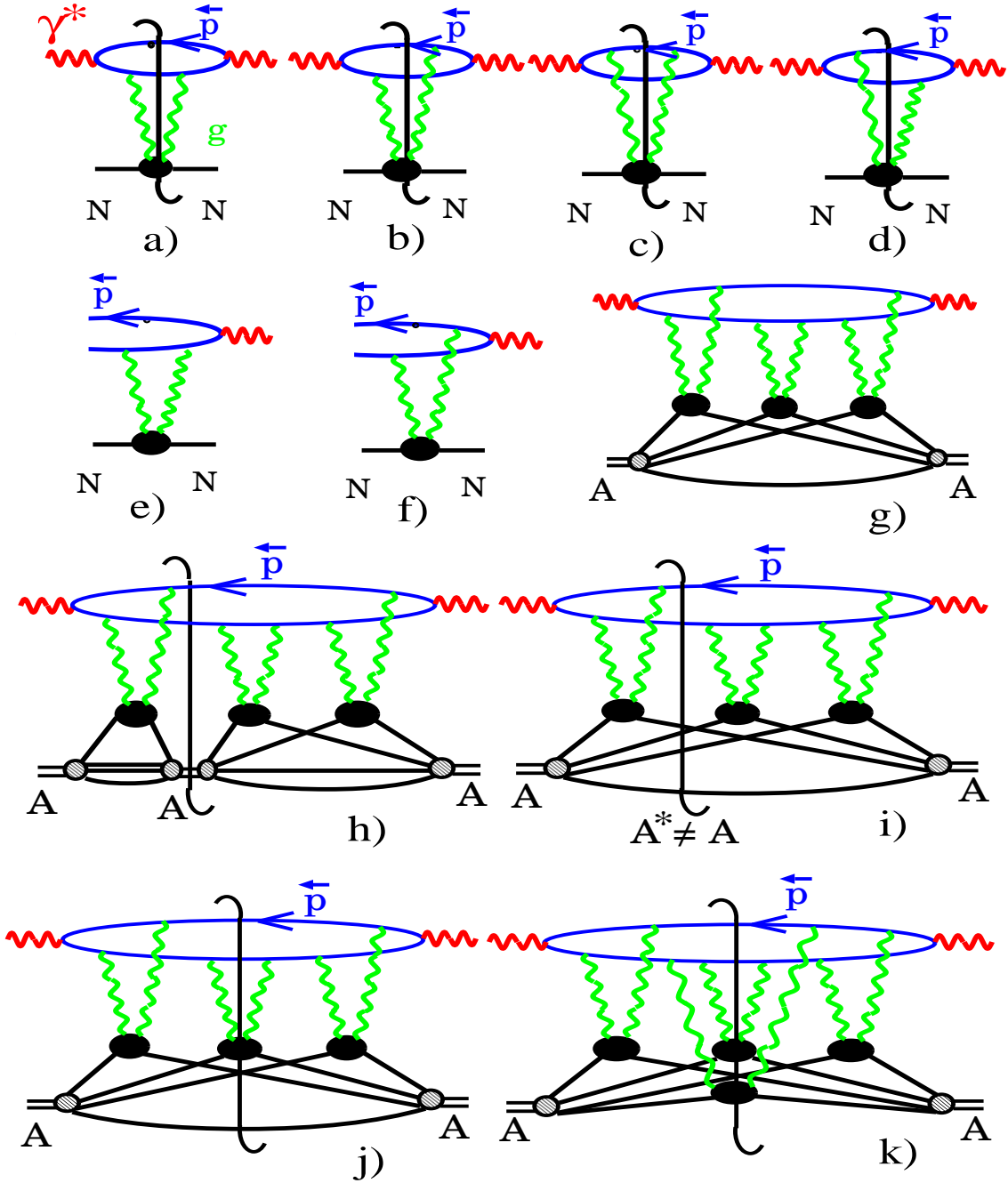, height=17.8cm, width = 15cm}
\end{center}
\caption{\it The pQCD diagrams for inclusive (a-d) and
diffractive
(e,f) DIS off protons and nuclei
(g-k). Diagrams (a-d) show the unitarity cuts with color
excitation of the target nucleon, (g) - a generic multiple
scattering diagram for Compton scattering off nucleus, (h) - the
unitarity cut for a coherent diffractive DIS, (i) - the unitarity
cut for quasielastic diffractive DIS with excitation of the
nucleus $A^*$, (j,k) - the unitarity cuts for truly inelastic DIS
with single and multiple color excitation of nucleons of the
nucleus. }
\end{figure}

For DIS off a free nucleon target, see figs. 3a-3d, the total
photoabsorption cross section equals \
\bea
\sigma_N = \int d^2\br dz |\Psi(z,\br)|^2 \sigma(\br)\, .
\label{eq:2.4} 
\eea
Upon the relevant Fourier transformations  one finds the 
momentum spectrum of the final state (FS) quark prior the hadronization,
\bea
{d\sigma_N \over d^2\bp dz} =
{\sigma_0\over 2}\cdot { \alpha_S(\bp^2) \over (2\pi)^2}
 \int d^2\bkappa f(\bkappa )
\left|\langle \gamma^*|\bp\rangle - \langle \gamma^*|\bp-\bkappa \rangle\right|^2
\label{eq:2.5}
\eea
where $\bp$ is the transverse momentum of the quark  and $z_+= z$ and $
z_- = 1-z$
are the fractions of photon's lightcone momentum carried by the quark 
and antiquark, respectively. The variables $z_{\pm}$ for the observed
jets add up to unity 
which is the signature
of the unresolved photon interaction.

For the reference, we cite here the relevant overlap of the momentum 
space wave functions of the photon, summed over the helicities of the
FS quark and antiquark \cite{NZ91}. For transverse photons
\bea
\left(\langle \gamma^*|\bp\rangle - \langle \gamma^*|\bp-\bkappa \rangle\right)
\!\!\!\!\!\!\!\! &&\cdot
\left.\left(\langle \bp |\gamma^*\rangle - \langle\bp-\bkappa | \gamma^*\rangle\right)
\right|_{\lambda_{\gamma}=\pm 1} = \nonumber\\
 2N_ce_f^2\alpha_{em}
\times \Bigg\{[z^{2}+(1-z)^{2}] \!\!\! \!\!\!\!\! &&\left.
\left({\bp \over  \bp^{2}+\varepsilon^{2}} - 
{\bp-\bkappa  \over  (\bp-\bkappa )^{2}+\varepsilon^{2}}\right)\right.
\nonumber\\
&& \!\!\! \cdot \left.\left({\bp \over  \bp^{2}+\varepsilon^{2}} - 
{\bp-\bkappa  \over  (\bp-\bkappa )^{2}+\varepsilon^{2}}\right)
\right|_{\lambda+\overline{\lambda}=0}\nonumber\\ 
 +m_{f}^{2} \!\!\!  \!\!\! \!\!\! &&
\left({1   \over  \bp^{2}+\varepsilon^{2}}-
{1 \over  (\bp-\bkappa )^{2}+\varepsilon^{2}}\right)\nonumber\\
\cdot\!\!\!\!\!\! \!\!\!\! &&  
\left.\left({1   \over  \bp^{2}+\varepsilon^{2}}-
{1 \over  (\bp-\bkappa )^{2}+\varepsilon^{2}}\right)
\right|_{\lambda+\overline{\lambda}=\lambda_{\gamma}}\Bigg\}
\label{eq:2.6}
\eea
and for the longitudinal photons
\bea
\left(\langle \gamma^*|\bp\rangle - \langle \gamma^*|\bp-\bkappa \rangle\right)\!\!\!\! \!\!\!\! &&\cdot
\left.\left(\langle \bp |\gamma^*\rangle - \langle\bp-\bkappa | \gamma^*\rangle\right)
\right|_{\lambda_{\gamma}=0} \nonumber\\
= 8N_c e_f^2\alpha_{em}\!\!\!\!\!\! \!\!\!\!&& Q^2z^2(1-z)^2 \nonumber\\
\times \left({1   \over  \bp^{2}+\varepsilon^{2}}-
{1 \over  (\bp-\bkappa )^{2}+\varepsilon^{2}}\right)\!\!\!\!\! \!\!\!\!
&&\cdot
\left. \left({1   \over  \bp^{2}+\varepsilon^{2}}-
{1 \over  (\bp-\bkappa )^{2}+\varepsilon^{2}}\right)
\right|_{\lambda+\overline{\lambda}=\lambda_{\gamma}}
\label{eq:2.7}
\eea
where
\beq
\varepsilon^2 = z(1-z)Q^2 + m_f^2
\label{eq:2.8}
\eeq
and $m_f$ is the mass of the quark $q_f$ with the charge $e_f$.
Now, notice that the transverse momentum of the gluon is precisely the
decorrelation momentum $\bDelta$, so that in the still further differential
from 
\bea
{d\sigma_N \over dz d^2\bp_+ d^2\bDelta} =
{\sigma_0\over 2}\cdot { \alpha_S(\bp^2) \over (2\pi)^2}
 f(\bDelta )
\left|\langle \gamma^*|\bp_+\rangle - 
\langle \gamma^*|\bp_+ -\bDelta \rangle\right|^2
\label{eq:2.9}
\eea
A useful small-$\Delta$ expansion for light flavors is 
\bea
{d\sigma_N \over dz d^2\bp_+ d^2\bDelta}& = &
{N_c e_f^2 \alpha_{em}\sigma_0 \alpha_S(\bp^2) \over (2\pi)^2}
\left[z^2 + (1-z)^2\right] \nonumber\\
& \times &  f(\bDelta ) 
\left\{ 
\left[{\varepsilon^2 -\bp_+^2 \over (\varepsilon^2 +\bp_+^2)^2}\right]^2
\Delta_L^2 + \left[{1\over \varepsilon^2 +\bp_+^2}\right]^2\Delta_{\perp}^2
\right\} 
\label{eq:2.10}
\eea
where $\Delta_{L}$ and $\Delta_{\perp}$ are the components of the 
decorrelation parallel, and transverse,
to $\bp_+$. Only the latter,
acoplanarity component, 
contributes to the azimuthal jet-jet decorrelation, the former is only
measurable if full experimental reconstruction of jets is possible.
This is the leading order result from the $k_{\perp}$ factorization, for
the applications to DIS off free nucleons see
\cite{Azimuth} and references therein. Below we shall extend 
(\ref{eq:2.9}) to jet-jet inclusive cross section
for interactions of unresolved photons with nuclear targets.

%-----------------  Section 3

\section{Propagation of color dipoles in nuclear me\-dium}
 
We focus on DIS at 
$x\lsim x_A = 1/R_A m_N \ll 1$ which is dominated by interactions 
of $q\bar{q}$ states of the photon. Extension to interactions
of higher Fock states of the photon is straightforward and
will give rise to the evolution effects, which we do not
consider here. For $x\lsim x_A$ the
propagation of the $q\bar{q}$ pair inside nucleus can be treated 
in the straight-path approximation.

We work in the conventional approximation of two t-channel gluons
in DIS off free nucleons. The relevant unitarity cuts of the
forward Compton scattering amplitude are shown in fig. 3a-3d 
and describe the transition from the color-neutral $q\bar{q}$
dipole to the color-octet $q\bar{q}$ pair. (Really, for $SU(N_c)$ 
we have fundamental \& adjoint multiplets instead of the
triplet \& octet familiar for $N_c=3$, our continuous 
reference to the latter should
not cause any confusion.) The two-gluon
exchange approximation amounts to neglecting small unitarity
corrections to DIS off free nucleons \cite{NZ94,BGNPZunit}, 
which are quantified
by diffractive DIS described by higher order diagrams of fig. 3e,3f.
This approximation 
is justified by a small fraction of diffractive DIS, $\eta_D \ll 1$
\cite{GNZ95}. The unitarity cuts of the nuclear Compton scattering
amplitude which correspond to the genuine inelastic DIS with color
excitation of the nucleus are shown in figs. 3j,3k. Here we notice
that in the diagram 3k the $q\bar{q}$ pair propagates 
in the both color-octet and color-singlet states.

Let $\bb_+$ and $\bb_-$ be the impact parameters of the quark
and antiquark, respectively, and $S_A(\bb_+,\bb_-)$  be the 
S-matrix for interaction of the $q\bar{q}$ pair with the nucleus.
We are interested in the inclusive cross section when we sum 
over all excitations of the target nucleus when one or several
nucleons have been color excited. A convenient way to sum such
cross sections is offered by the closure relation. Regarding 
the color states $c_{km}$ of the $q_k\bar{q}_m$ pair, we sum over all octet
and singlet states. Then, the the 2-body inclusive
spectrum is calculated in terms of the 2-body density matrix as 
\bea
&&{d\sigma_{in} \over dz d^2\bp_+ d^2\bp_-} =
{1\over (2\pi)^4} \int d^2 \bb_+' d^2\bb_-' d^2\bb_+ d^2\bb_- \nonumber\\
&&\times \exp[-i\bp_+(\bb_+ -\bb_+')-i\bp_-(\bb_- -
\bb_-')]\nonumber\\
&&\times \Psi^*(\bb_+' -\bb_-')
\Psi(\bb_+ -\bb_-)\nonumber\\
&&\times \left\{\sum_{A^*} \sum_{km}  \langle
1;A|S_A^*(\bb_+',\bb_-')|A^*;c_{km}\rangle
\langle
c_{km};A^*|S_A(\bb_+,\bb_-)|A;1\rangle \right. \nonumber\\
&& - 
\left. \langle 1;A|S_A^*(\bb_+',\bb_-')
|A;1\rangle
\langle 1;A|S_A(\bb_+,\bb_-)|A;1\rangle \right\} 
\label{eq:3.1} 
\eea
Here $\Psi$ is the $q \bar{q}$--Fock state wave function of the
virtual photon, and we suppressed its dependence on $z_{\pm}$. In the
integrand of (\ref{eq:3.1}) we subtracted the diffractive component
of the final state. Notice, that the four straight-path trajectories 
$\bb_{\pm},\bb'_{\pm}$ enter the calculation of the full fledged
2-body density matrix  and $S_A$ and $S_A^*$ describe the propagation of
two $q\bar{q}$ pairs inside a nucleus. 

Upon the application of closure to sum over nuclear final states $A^*$ 
the integrand of (\ref{eq:3.1}) can be considered as an intranuclear
evolution operator for the 2-body density matrix (for the related
discussion see ref. \cite{NSZdist}) 
\bea
\sum_{A^*} \sum_{km} \langle A| \left\{
\langle 1| S_A^*(\bb_+',\bb_-')|c_{km}\rangle \right\}
|A^* \rangle
\langle A^*| \left\{ \langle c_{km}| S_A(\bb_+,\bb_-)|1\rangle \right\}
|A\rangle =\nonumber\\
=\langle
A| \left\{ \sum_{km} \langle 1| S_A^*(\bb_+',\bb_-')|c_{km}\rangle
\langle
c_{km}|S_A(\bb_+,\bb_-)|1\rangle\right\} |A\rangle 
\label{eq:3.2}
\eea
Let the QCD eikonal for the quark-nucleon and antiquark-nucleon
one-gluon exchange interaction
be $T^a_{+} \Delta(\bb)$ and  $T^a_{-} \Delta(\bb)$,
where  $T^{a}_+$ and $T^{a}_{-}$ are the 
$SU(N_c)$ generators for the quark and antiquark states, 
respectively. The vertex 
$V_a$ for excitation of the nucleon $g^a N \to N^*_a$ into
color octet state is so normalized that after application 
of closure the vertex $g^a g^b NN$ 
in the diagrams of fig. 3a-3d is $\delta_{ab}$. Then, to the 
two-gluon exchange approximation, the $S$-matrix of the 
$(q\bar{q})$-nucleon interaction equals
\bea
S_N(\bb_+,\bb_-)  = 1 + i[T^a_{+} \Delta(\bb_+)+ 
T^a_{-} \Delta(\bb_-)]V_a
- {1\over 2} [T^a_{+} \Delta(\bb_+)+ T^a_{-} \Delta(\bb_-)]^2
\label{eq:3.3}
\eea
and the color-dipole cross section for the $q\bar{q}$ dipole
is
\bea
\sigma(\bb_+ - \bb_-) = {N_c^2 -1 \over 2 N_c} \int d^2\bb_+
[\Delta(\bb_+)-\Delta(\bb_-)]^2\, .
\label{eq:3.4}
\eea

The nuclear $S$-matrix of the straight-path approximation is
given by 
\bea
S_A(\bb_+,\bb_-) = \prod _{j=1}^A S_N(\bb_+ -\bb_j,\bb_- - \bb_j) 
\label{eq:3.5}
\eea
where $\bb_{j}$ are the transverse coordinates of nucleons in
a nucleus and the ordering along the longitudinal path is understood.
We evaluate the nuclear expectation value in (\ref{eq:3.2})
in the standard dilute gas approximation.
To the two-gluon exchange approximation, only the terms quadratic
in $\Delta(\bb_j)$ must be kept in the evaluation of the
single-nucleon matrix element 
$$
\langle N_j|S_N^*(\bb_+' -\bb_j,\bb_-' - \bb_j)
S_N(\bb_+-\bb_j,\bb_- - \bb_j)|N_j \rangle
$$
per each and every nucleon which enters 
the calculation of $S_A^*S_A$. 
Following the technique developed in \cite{NPZcharm,LPM} we can
reduce the calculation of the evolution operator for the 
2-body density matrix (\ref{eq:3.2})
to the evaluation of the $S$-matrix $S_{4A}(\bb_+,\bb_-,\bb_+',\bb_-')$
for a scattering of a fictitious 4-parton state composed of the 
two quark-antiquark pairs in the overall color-singlet state. 
Because $(T^{a}_+)^*= -T^{a}_{-}$, within the two-gluon 
exchange approximation the quarks entering the complex-conjugate 
$S_A^*$ in (\ref{eq:3.2}) can be viewed as antiquarks, so that 
\bea
\langle
\left\{ \sum_{km} \langle 1| S_A^*(\bb_+',\bb_-')|c_{km}\rangle
\langle
c_{km}|S_A(\bb_+,\bb_-)|1\rangle\right\}  
=\nonumber\\
=\sum_{km jl} \delta_{kl} \delta_{mj}
\langle c_{km}c_{jl}|S_{4A}(\bb_+',\bb_-',\bb_+,\bb_-)|11\rangle 
\label{eq:3.6}
\eea
where $S_{4A}(\bb_+',\bb_-',\bb_+,\bb_-)$ is an
S-matrix for the propagation of the two quark-antiquark pairs in the
overall singlet state. While the first $q\bar{q}$ pair is formed 
by the initial quark $q$ and antiquark $\bar{q}$ at impact parameters 
$\bb_{+}$ and $\bb_-$, respectively, in the second pair $q'\bar{q}'$ 
the quark $q'$ propagates at an impact parameter $\bb_-'$ and the
antiquark $\bar{q}'$ at an impact parameter $\bb_+'$, as indicated 
in fig.~4.  
In the initial state the both quark-antiquark pairs are in
color-singlet states: $ | in \rangle = |11\rangle$.

%--------------   figure 4

\begin{figure}[!htb]
   \centering
   \epsfig{file=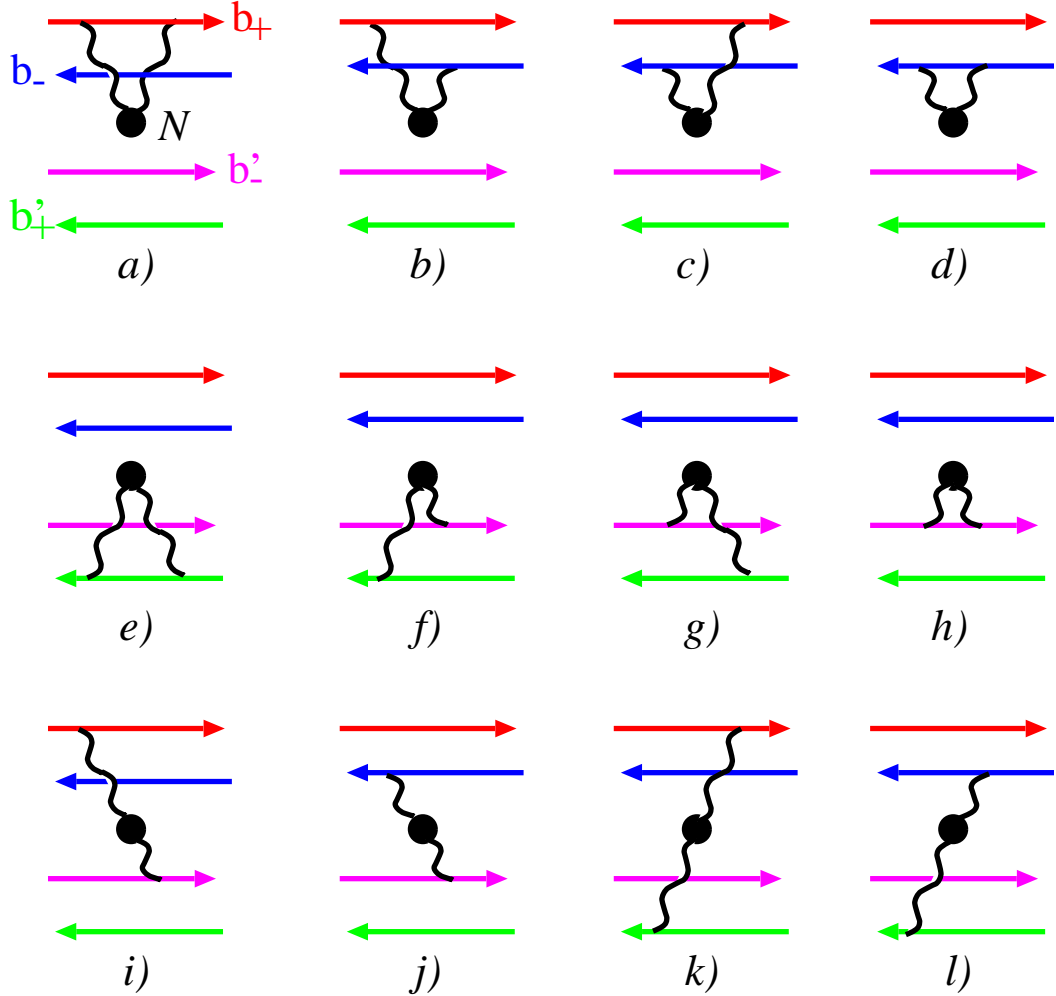,width=14cm}
\caption{\it The pQCD diagrams for the matrix of color dipole
cross section for the 
4-body $(q\bar{q})(q'\bar{q}')$ state. The set 4i-4l shows only half of
the diagrams for scattering with rotation of the color state of dipoles.}
\end{figure}

Let us introduce the normalized singlet-singlet and octet-octet
states
\beq
|11\rangle = {1\over N_c} (\bar{q} q)(\bar{q}' q')
\label{eq:3.7} 
\eeq
\beq
|88\rangle = {2\over \sqrt{N_c^2 -1}} (\bar{q}T^a q)(\bar{q}' T^a q') 
\label{eq:3.8} 
\eeq
where $N_c$ is the number of colors and $T^{a}$ are the generators
of $SU(N_c)$. Making use of the color Fiertz identity, 
\beq
\delta_{km} \delta_{jl} = {1\over N_c} \delta^k_l \delta^m_j +
2 \sum _{a} (T^{a})^k_l(T^{a})^m_j
\label{eq:3.9}
\eeq
the sum over
color states of the produced quark-antiquark pair can be 
represented as
\bea
\sum_{km}
\langle c_{km} c_{km}|S_{4A}(\bb_+',\bb_-',\bb_+,\bb_-)|11\rangle =
\langle 11|S_{4A}(\bb_+',\bb_-',\bb_+,\bb_-)|11\rangle \nonumber\\
+ \sqrt{N_c^2 -1} \langle 88|S_{4A}(\bb_+',\bb_-',\bb_+,\bb_-)|11\rangle 
\label{eq:3.10} 
\eea

If $\sigma_4(\bb_+',\bb_-',\bb_+,\bb_-)$ is the color-dipole
cross section operator for the 4-body state, then the standard
evaluation of the nuclear expectation value for a 
dilute gas nucleus gives
\bea
S_{4A}(\bb_+',\bb_-',\bb_+,\bb_-)=\exp\{- {1\over
2}\sigma_{4}(\bb_+',\bb_-',\bb_+,\bb_-)T(\bb)\}
\label{eq:3.11} 
\eea
where
\bea
T(\bb)=\int dz n_{A}(z, \bb)
\label{eq:3.12} 
\eea
is the optical thickness of a
nucleus at an impact parameter $\bb \approx \bb_{\pm},\bb_{\pm}'$.
In writing down (\ref{eq:3.6}) we employed the standard 
approximation of neglecting the size of color dipoles
compared to a radius of heavy nucleus.
The single-nucleon $S$-matrix (\ref{eq:3.3}) contains transitions
from the color-singlet to the both color-singlet and color-octet
$q\bar{q}$ pairs. However, only the color-singlet operators
contribute to 
$\langle N_j|S_N^*(\bb_+' -\bb_j,\bb_-' - \bb_j)
S_N(\bb_+-\bb_j,\bb_- - \bb_j)|N_j \rangle$, and the matrix 
$\sigma_4(\bb_+',\bb_-',\bb_+,\bb_-)$ only includes transitions
between the $|11\rangle$ and $|88\rangle$ color-singlet 4-parton
states, the $|18\rangle$ states are not allowed. 

 Performing the relevant color
algebra, we find
\bea
\langle 11|\sigma_4|11\rangle = \sigma(\bb_+ - \bb_-)+\sigma(\bb_+' - \bb_-')
\label{eq:3.13} 
\eea
\bea
\langle 11|\sigma_4|88\rangle = 
{1\over \sqrt{N_c^2-1}}
[&&\sigma(\bb_- - \bb_-')+\sigma(\bb_+ - \bb_+')\nonumber\\
-
 &&\sigma(\bb_+ - \bb_-')- \sigma(\bb_-  - \bb_+')]
\label{eq:3.14} 
\eea
\bea
\langle 88|\sigma_4|88\rangle = 
&&{N_c^2 -2 \over N_c^2-1}[
[\sigma(\bb_+ - \bb_+')+\sigma(\bb_- - \bb_-')]\nonumber\\
+&&{2 \over N_c^2-1}[\sigma(\bb_+ - \bb_-')+ \sigma(\bb_-  - \bb_+')]\nonumber\\ 
-&&{1\over N_c^2-1}
[\sigma(\bb_+ - \bb_-)+\sigma(\bb_+' - \bb_-')]
\label{eq:3.15} 
\eea
The term in (\ref{eq:3.1}), which subtracts the contribution from processes 
without color excitation of the target nucleus, equals
\bea  
\langle A| \left\{\langle 1|S_A^*(\bb_+',\bb_-')
|1\rangle\right\}|A\rangle
\langle A|\left\{
\langle 1|
S_A(\bb_+,\bb_-)|1\rangle\right\}|A\rangle \nonumber\\
= \exp\{- {1\over
2}\left[\sigma(\bb_+ - \bb_-)+\sigma(\bb_+' - \bb_-')\right]T(\bb)\}
 \label{eq:3.17} 
\eea

Let $\Sigma_{1,2}$ be the two eigenvalues of the operator $\sigma_4$. Then
it is convenient to use the Sylvester expansion
\bea
&&\exp\{- {1\over
2}\sigma_4 T(\bb)\} = \nonumber\\
&&\exp\{- {1\over
2}\Sigma_1 T(\bb)\}{\sigma_4 - \Sigma_2 \over \Sigma_1 - \Sigma_2}
+  \exp\{- {1\over
2}\Sigma_2 T(\bb)\}{\sigma_4 - \Sigma_1 \over \Sigma_2 - \Sigma_1}
\label{eq:3.18} 
\eea
Evidently, in the general case the eigen-cross sections are fairly complex
functions of the six possible dipoles formed by four partons. One
may wonder how do the standard color-dipole Glauber-Gribov formulas
\cite{NZ91,NZZdiffr} for nuclear DIS cross sections emerge from (\ref{eq:3.18})?   
Hence we turn first to the simpler cases of the total inelastic 
and single-jet inclusive cross sections.
An application to (\ref{eq:3.10}) of the Sylvester expansion gives
\bea
&&(\langle 11| + \sqrt{N_c^2 -1} \langle 88|)
\exp\left\{-{1\over 2}\sigma_4 T(\bb)\right\}|11\rangle\nonumber\\
&& -
\exp\left\{-{1\over 2}\left[\sigma(\br)+\sigma(\br')\right]
T(\bb)\right\}\nonumber\\
&&=
\exp\left\{-{1\over 2}\Sigma_2 T(\bb)\right\}-
\exp\left\{-{1\over 2}\left[\sigma(\br)+\sigma(\br')\right]
T(\bb)\right\}\nonumber\\
&&+ { \langle 11| \sigma_4 | 11\rangle - \Sigma_2 \over \Sigma_1 - \Sigma_2}
\left\{\exp\left[-{1\over 2}\Sigma_1 T(\bb)\right]-
\exp\left[-{1\over 2}\Sigma_2 T(\bb)\right]\right\}\nonumber\\
&&+{\sqrt{N_c^2 -1} \langle 11| \sigma_4 | 88\rangle \over \Sigma_1 - \Sigma_2}
\left\{\exp\left[-{1\over 2}\Sigma_1 T(\bb)\right]-
\exp\left[-{1\over 2}\Sigma_2 T(\bb)\right]\right\}
\label{eq:3.19}
\eea

% -------------  Section 4

\section{Non-Abelian aspects of propagation of color dipoles 
in nuclear medium and Glauber-Gri\-bov formalism}

In order to get an insight into the impact of propagation
of the color-octet $q\bar{q}$ pairs inside a nucleus let us 
consider first the total inelastic cross section obtained from
(\ref{eq:3.1}) upon the integration over the transverse momenta
$\bp_{\pm}$  of the quark and antiquark. Such an integration amounts
to putting $\bb_+=\bb_+'$ and $\bb_-=\bb_-'$. Then we are left with 
the system of two color dipoles of the same size $\br=\bb_+ -\bb_- =
\br'=\bb_+' -\bb_-'$, 
and the matrix of the 4-body cross sections has the eigenvalues
\bea
\Sigma_1 = 0\, ,
\label{eq:4.1}
\eea
\bea
\Sigma_2 = {2N_c^2 \over N_c^2 -1} \sigma(\br)
\label{eq:4.2} 
\eea
with the eigenstates 
\bea
|f_1\rangle = {1\over N_c}(|11\rangle + \sqrt{N_c^2-1} |88\rangle)\, ,
\label{eq:4.3} 
\eea
\bea
|f_2\rangle = {1\over N_c}(\sqrt{N_c^2-1}|11\rangle - |88\rangle)\, .
\label{eq:4.4} 
\eea
The eigen-cross section (\ref{eq:4.2}) differs from $\sigma(\br)$  
for the color-singlet $q\bar{q}$ pair by the nontrivial 
color factor $N_c^2/(N_c^2-1)$. The
existence of the non-attenuating 4-quark state with $\Sigma_1=0$
is quite obvious and corresponds to an overlap of two dipoles
of the same size with neutralization of color charges.

Now notice, that the final state which enters the calculation of
the genuine inelastic DIS off a nucleus, see eq. (\ref{eq:3.10}), 
is precisely the eigenstate $|f_1\rangle$.
Then, even without invoking the Sylvester expansion (\ref{eq:3.18}),
(\ref{eq:3.19}), the straightforward result for  the inelastic 
cross section is
\bea
\sigma_{in} = &&\int d^2\br dz |\Psi(z,\br)|^2 \nonumber\\
\times &&\int d^2\bb
\left\{ N_c\langle f_1| \exp\left[-{1\over 2}\sigma_4 T(\bb)\right]
|11\rangle
-\exp\left[-\sigma(\br) T(\bb)\right]\right\} \nonumber\\
= &&\int d^2\bb\langle \gamma^*| 
\left\{\exp\left[-{1\over 2}\Sigma_1 T(\bb)\right]
-\exp\left[-\sigma(\br) T(\bb)\right]\right\} |\gamma^* \rangle \nonumber\\
= &&\int d^2\bb\langle \gamma^*|
\left\{1
-\exp\left[-\sigma(\br) T(\bb)\right]\right\}|\gamma^* \rangle  
\label{eq:4.5} 
\eea
what is precisely the Glauber-Gribov formula \cite{NZZdiffr} 
in which no trace of
a propagation inside a nucleus of the eigenstate (\ref{eq:4.4}) with
the eigen-cross section (\ref{eq:4.2}) is left.

When the final state $q\bar{q}$ is produced in the color-singlet
state, the net flow of color between the $q\bar{q}$ pair and 
color-excited debris of the target nucleus is zero. Which suggests
that a rapidity gap may survive upon the hadronization, although 
whether a rapidity gap in genuine inelastic events with color-singlet 
$q\bar{q}$ production is stable against higher order correction or 
not remains an interesting open issue. Here we proceed under the
assumption that the rapidity gap survives and the production of
the color-singlet and color-octet $q\bar{q}$ pairs can be
separated. Although
the debris of the target nucleus have a zero net color charge,
the color-excited debris of nucleons are spatially separated by a 
distance of the order of the nuclear radius. The formation of
color strings between the color centers would lead to a total
excitation energy of the order of 1 GeV times $A^{1/3}$, so that
such a rapidity-gap events would look like a double diffraction with
multiple production of mesons in the nucleus fragmentation region
(for the theoretical discussion of conventional mechanisms of 
diffraction
excitation of nuclei in proton-nucleus collisions see 
\cite{ZollerDiffrNucl}, the experimental observation has
been reported in \cite{HELIOSNuclDiffr}).  
As such it is distinguishable from quasielastic diffractive DIS
followed by excitation and breakup of the target nucleus without
production of secondary particles. 

Making use of the Sylvester expansion (\ref{eq:3.18})-(\ref{eq:3.19}), 
and the
eigenstates (\ref{eq:4.3}),(\ref{eq:4.4}), one readily obtains
\bea
\sigma_{in}(A^*(q\bar{q})_1) = ~~~~~~~~~~~~~~~~~~~~~~~~~~~~~
\nonumber\\\int d^2\bb 
\langle \gamma^*|\left\{ (1
-\exp\left[-\sigma(\br) T(\bb)\right]) -
{N_c^2-1 \over N_c^2} (1
-\exp\left[-{1\over 2}\Sigma_2 T(\bb)\right])\right\}|\gamma^* \rangle\,  
\label{eq:4.6} 
\eea
and
\bea
\sigma_{in}(A^*(q\bar{q})_8) = {N_c^2-1 \over N_c^2} \int d^2\bb
\langle \gamma^*|\left\{1
-\exp\left[-{1\over 2}\Sigma_2 T(\bb)\right]\right\}|\gamma^* \rangle \, .
\label{eq:4.7} 
\eea
In the latter case the nontrivial 
eigen-cross section $\Sigma_2$ enters explicitly.

Several features of the result (\ref{eq:4.6}) are noteworthy.
First, the color neutralization of the $q\bar{q}$ pair after
the first inelastic interaction requires at least one secondary
inelastic interaction. Indeed, an expansion of the integrand
starts with the term quadratic in the optical thickness:
\bea
&&\left\{(1
-\exp\left[-\sigma(\br) T(\bb)\right]) -
{N_c^2-1 \over N_c^2} (1
-\exp\left[-{1\over 2}\Sigma_2 T(\bb)\right])\right\} \nonumber\\
&& = {\sigma^2(\br) T^2(\bb) \over 2 (N_c^2-1)}+...
\label{eq:4.8} 
\eea
Second, in the large-$N_c$ limit the color octet state tends to
oscillate in color remaining in the octet state. This is clearly
seen from (\ref{eq:4.8}). Third, in the limit of an opaque
nucleus 
\bea
\sigma_{in}(A^*(q\bar{q})_1) = {1 \over N_c^2}\int d^2\bb
\langle \gamma^*|\left\{1
-\exp\left[-\sigma(\br) T(\bb)\right]\right\}|\gamma^* \rangle 
= {1\over N_c^2} \sigma_{in}
\label{eq:4.9} 
\eea
and remains a constant fraction of DIS in contrast to the
quasielastic diffractive DIS or inelastic diffractive excitation
of a nucleus, the cross sections of which 
vanish for an opaque nucleus \cite{NZZdiffr,ZollerDiffrNucl}.

The result (\ref{eq:4.7}) shows how the non-Abelian nature
of the propagation of color dipoles inside a nucleus which
manifests itself via the eigen-cross section (\ref{eq:4.2})
distinct from $\sigma(\br)$ reveals itself in the total
cross section of inelastic DIS with excitation of dijets 
in the color-octet state.

An analysis of the single-parton, alias the single-jet, 
inclusive cross section is
quite similar. In this case we integrate over the momentum
$\bp_-$ of the antiquark jet, so that $\bb_-'=\bb_-$. The
matrix $\sigma_4$ has the eigenvalues
\bea
\Sigma_1 = \sigma(\br-\br')
\label{eq:4.10}
\eea
\bea
\Sigma_2 = {N_c^2 \over N_c^2-1} \left[\sigma(\br)+\sigma(\br')\right]
-{1\over N_c^2-1}\sigma(\br-\br')
\label{eq:4.11} 
\eea
with exactly the same eigenstates $|f_1\rangle$ and $|f_2\rangle$ as
given by eqs. (\ref{eq:4.3}),(\ref{eq:4.4}). Again, the cross section
of genuine inelastic DIS corresponds to projection onto the eigenstate
$|f_1\rangle$, so that \cite{Saturation}
\bea {d \sigma_{in}\over d^2\bb d^2\bp dz }   &=&  {1
\over (2\pi)^2}
 \int d^2\br' d^2\br
\exp[i\bp(\br'-\br)]\Psi^*(\br')\Psi(\br)\nonumber\\
&\times& \left\{\exp[-{1\over 2}\Sigma_1 T(\bb)]-
\exp[-{1\over 2}[\sigma(\br)+\sigma(\br')]T(\bb)]\right\}\nonumber\\
&=&  {1
\over (2\pi)^2}
 \int d^2\br' d^2\br
\exp[i\bp(\br'-\br)]\Psi^*(\br')\Psi(\br)\nonumber\\
&\times& \left\{\exp[-{1\over 2}\sigma(\br-\br') T(\bb)]-
\exp[-{1\over 2}[\sigma(\br)+\sigma(\br')]T(\bb)]\right\}
\, .
\label{eq:4.12} 
\eea

% ------------------ Section 5

\section{The Pomeron-Splitting Mechanism for
Dif\-fractive Hard Dijets and Weizs\"acker-Wil\-l\-i\-ams glue of nuclei}

At QCD parton level  diffraction dissociation of photons and          
hadrons is modeled by excitation of $q\bar{q}, q\bar{q}g, ...$         
Fock states which are lifted on their mass shell through the          
$t$--channel exchange of a QCD Pomeron with the target hadron.          
The color singlet two--gluon structure of the Pomeron gives rise         
to the two distinct forward $q\bar{q}$ dijet production         
subprocesses: The first one, of fig.~3e (also fig. 5a), 
is a counterpart of          
the classic Landau-Pomeranchuk         
\cite{Landau,Feinberg,Glauber} mechanism of diffraction dissociation          
of deuterons into the proton--neutron continuum and can be dubbed  
the splitting of the beam particle into the dijet, because the transverse          
momentum $\bk$ of jets comes from the intrinsic transverse momentum          
of quarks and antiquarks in the beam particle.  
Specific of QCD is the mechanism of fig.~3f (also fig.~5b) where jets receive          
a transverse momentum from gluons in the Pomeron.         
In an extension of the early work \cite{NZ92}, in \cite{NZsplit}
it was shown  that the second         
mechanism dominates at a sufficiently large $\bk$ and in this          
regime diffractive amplitudes are proportional to the differential          
(unintegrated) gluon structure function ${\cal{F}}(x,\bk^2) =          
\partial G(x, \bk^2)/\partial \log \bk^2$. Correspondingly, this         
mechanism has been dubbed splitting of the Pomeron into dijets.         
In diffractive DIS the Pomeron splitting dominates at $k\gg Q$,         
whereas the somewhat modified Landau et al. mechanism dominates         
at $k \lsim Q$.         
         
Motivated by the recent data from the E791 Fermilab         
experiment \cite{Ashery}, here we discuss         
peculiarities of the Pomeron splitting mechanism on an
example of diffractive         
excitation of pions into dijets on free nucleon and nuclear targets.  
In a slight adaption         
of the results of \cite{NZ92,NZsplit} one should replace the          
pointlike $\gamma^* q\bar{q}$ vertex $eA_\mu \bar{\Psi}\gamma_\mu\Psi$         
by the non--pointlike $\pi q\bar{q}$ vertex $i \Gamma(M^2) \bar{\Psi}         
\gamma_5 \Psi$. Here $M^2 = (\bk^2+m_f^2)/z(1-z)$ is the invariant         
mass squared of the dijet system, the pion momentum is shared by the jets         
in the partitioning $z,1-z$.          
The explicit form of the spinor vertex reads     
\begin{equation}         
\overline{\Psi}_{\lambda}(\bk)\gamma_{5}\Psi_{\bar{\lambda}}(-\bk)          
= {\lambda \over \sqrt{z(1-z)}}         
[m_{f} \delta_{\lambda -\bar{\lambda}} - \sqrt{2}\bk\cdot \bfe_{-\lambda}         
\delta_{\lambda \bar{\lambda}}]\, ,         
\label{eq:5.1}         
\end{equation}         
where $m_f$ is the quark mass, $\lambda$ and $\bar{\lambda}$ are the quark         
and antiquark helicities and $\bfe_\lambda = -(\lambda \bfe_x +i \bfe_y )/\sqrt{2}$.         
The two helicity amplitudes $\Phi_0(z,\bk,\bDelta)$ for          
$\lambda+\bar{\lambda}=0$ and          
$\bPhi_1(z,\bk,\bDelta)$ for $\lambda +\bar{\lambda}=\pm 1$,          
can be cast in the form (we concentrate          
on the forward limit, $\bDelta =0$)         
\begin{eqnarray}         
\Phi_0(z,\bk)=\alpha_{S}(\bk^2)\sigma_{0}\left[         
\int d^2\bkappa m_{f}\psi_{\pi} (z,\bk) f(\bkappa)         
- \int d^2\bkappa         
m_{f}\psi_{\pi} (z,\bkappa)f(\bk-\bkappa)\right]         
\, ,         
\label{eq:5.2}         
\end{eqnarray}          
and $\bPhi_1$ is obtained from the substitution          
$m_f \psi(z,\bk) \to \bk \psi(z,\bk)$.         
The radial wave function of the $q\bar{q}$         
Fock state of the pion is related to the $\pi q\bar{q}$ vertex function as  
\begin{equation}         
\psi_{\pi}(z,\bk)={N_{c}\over 4\pi^3 z(1-z)}         
{\Gamma_{\pi}(M^{2})\over ( M^2 -m_{\pi}^2)} \,          
\label{eq:5.3}         
\end{equation}         
and is normalized to the $\pi \to \mu \nu$ decay constant $F_\pi = 131$ MeV through         
\begin{equation}         
F_\pi = \int d^2\bk dz m_{f} \psi_{\pi}(z,\bk)         
=F_{\pi}\int_0^1 dz \phi_{\pi}(z)\, ,         
\label{eq:5.4}         
\end{equation}         
where $\phi_\pi(z)$ is the pion distribution amplitude \cite{PionDA}.      
Finally, our normalization of helicity         
amplitudes is such, that the differential cross section of forward          
dijet production equals         
\begin{equation}         
\left.{d\sigma_{D} \over dz d\bk^2 d\bDelta^2}\right|_{\bDelta=0}         
= {\pi^3 \over 24} \left\{|\Phi_0|^2 + |\bPhi_1|^2 \right\} \, .   
\label{eq:5.5}       
\end{equation}         
         
We turn to the discussion of the asymptotics for large jet momenta $\bk$.         
 The first term in eq.(\ref{eq:5.2}) comes from the Landau et al. pion         
splitting mechanism of fig.~5a, whereas the second one is the         
contribution from the Pomeron splitting of fig.~5b. Because $\psi_{\pi}\otimes f$
is a convolution of nonoscillating functions, it         
is necessarily a broader function than $\psi_\pi(z,\bk)$ and thus         
will take over if only $\bk$ is large enough. For the quantitative          
estimate, we remind the reader, that for $x \sim 10^{-2}$ relevant to the  
kinematics of E791, the          
large--$\bk$ behavior of $f(\bk)$ is well described by the inverse          
power law  $f(\bk) \propto k^{-2\delta}$ with         
an exponent $\delta \sim 2.15$ \cite{INDiffGlue}.         
Clearly, $f(\bk)$ decreases much slower than $\psi(z,\bk)$, and hence         
the asymptotics of the convolution integral is controlled          
by the  asymptotics of $f(\bk)$:         
\begin{equation}         
\int d^2\bkappa         
m_{f}\psi_{\pi} (z,\bkappa)f(\bk-\bkappa) \approx f(\bk)\int d^2\bkappa         
m_{f}\psi_{\pi} (z,\bkappa) = f(\bk) \, \phi_{\pi}(z)\, F_\pi  \, ,
\label{eq:5.6}        
\end{equation}         
which shows that in this regime the dijet momentum comes entirely from the         
momentum of gluons in the Pomeron. Furthermore, in the same regime diffraction          
into dijets probes the pion distribution amplitude $\phi_{\pi}(z)$.

.
%----------------- figure 5
 
\begin{figure}[!htb]         
\begin{center}         
\epsfig{file=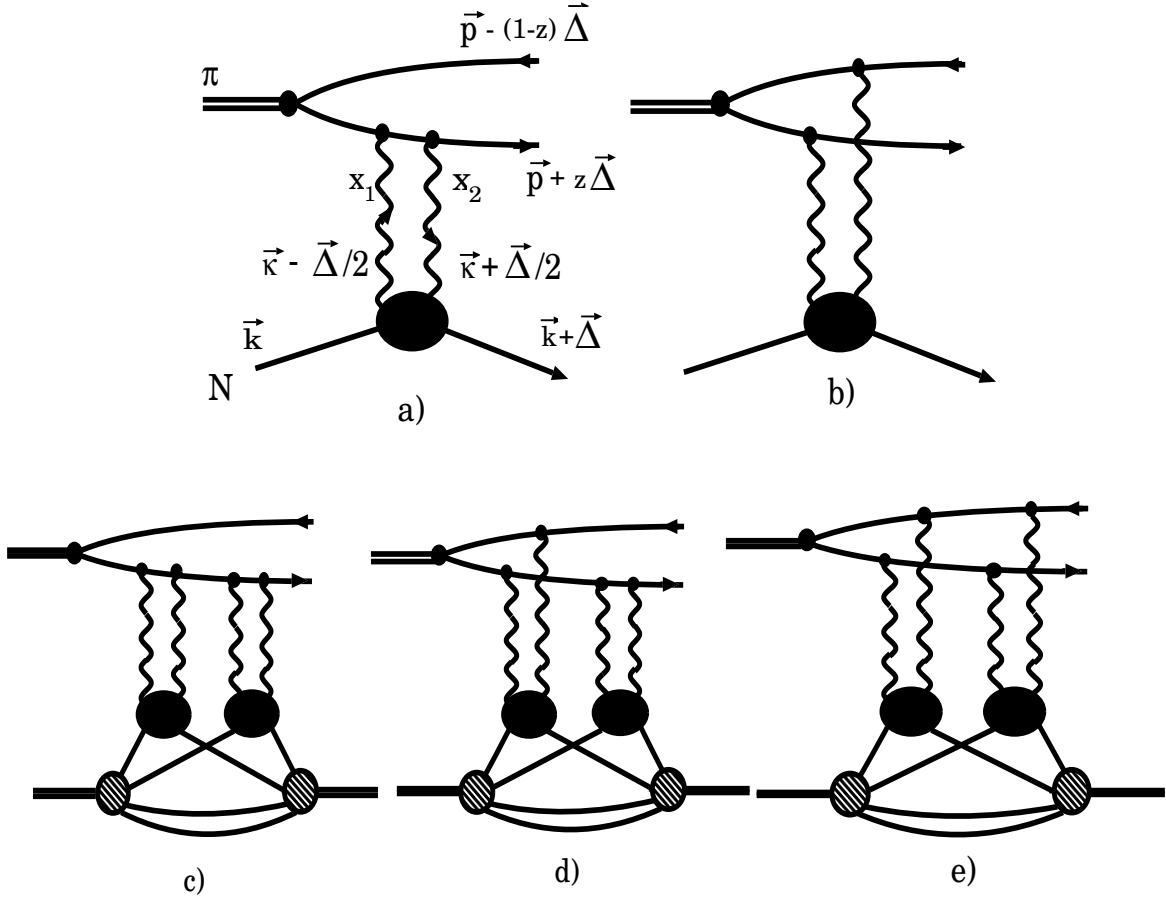, height = 15.5cm, width=12.0cm,angle=270}         
\end {center}         
\caption{\it Sample Feynman diagrams for diffractive dijet excitation  
in $\pi N$ collisions [diagrams 5a),5b)] and typical rescattering 
corrections to the nuclear coherent amplitude [diagrams 5c),5d),5e)].}         
\label{fig5}         
\end{figure}

Now we notice that in the color dipole representation
\begin{equation}         
\Phi_0 (z,\bk) = \, \int d^2\br          
\, e^{\displaystyle -i\bk \br} \, \sigma(\br)         
\, m_{f}\Psi_\pi(z, \br) \, ,         
\label{eq:5.7}         
\end{equation}         
and          
the nuclear amplitude is readily obtained \cite{NZ91} by substituting in         
eq.(\ref{eq:5.7}) 
\beq
\sigma(\br) \to \sigma_A(\br) =         
2 \, \int d^2\bb \left\{1-\exp[-{1\over 2} \sigma(\br) T_A(\bb)]\right\}
\label{eq:5.8}
\eeq
Typical nuclear double scattering diagrams of figs.~5c-5e can         
conveniently be classified as  nuclear shadowing of the pion splitting (fig. 5c),         
nuclear shadowing of single Pomeron splitting (fig.~5d) and double Pomeron          
splitting (fig.~5e) contributions. The $j$ Pomeron splitting         
is due to diagrams, in which $j$ of the Pomerons exchanged between the color          
dipole and the nucleus couple with one gluon to the quark and with one gluon          
to the antiquark and involve the $j$-fold convolutions of the unintegrated          
gluon distribution         
\beq 
f^{(j)}(\bkappa )= \int \prod_{i=1}^j
d^2\bkappa _{i} f(\bkappa _{i}) \delta(\bkappa -\sum_{i=1}^j
\bkappa _i) \, .
\label{eq:5.9} 
\eeq 
Now we can invoke the NSS representation for the nuclear attenuation 
factor \cite{NSSdijet,Saturation}
\bea
\exp\left[-{1\over 2}\sigma(\bs) T(\bb)\right] = 
&&\exp\left[-\nu_{A}(\bb)\right] 
\exp\left[\nu_{A}(\bb)\int 
d^2\bkappa f(\bkappa)\exp(i\bkappa\bs)\right]\nonumber\\
=&&\exp\left[-\nu_{A}(\bb)\right]
\sum_{j=0}^{\infty} {\nu_{A}^{j}(\bb) \over j!}\int d^2\bkappa 
f^{(j)}(\bkappa)\exp(i\bkappa\bs)\nonumber\\
 = &&\int d^2\bkappa \Phi(\nu_{A}(\bb), \bkappa)
\exp(i\bkappa\bs)
\label{eq:5.10}
\eea
where 
\bea
\Phi(\nu_{A}(\bb), \bkappa) = \exp(-\nu_{A}(\bb))f^{(0)}(\bkappa)
+\phi_{WW}(\bb,\bkappa)\, ,
\label{eq:5.11}
\eea
\bea
\nu_{A}(\bb)= {1\over 2}\alpha_S(r)\sigma_0 T(\bb) 
\label{eq:5.12}
\eea
and it is understood that  
\bea
f^{(0)}(\bkappa)=\delta(\bkappa)\, .
\label{eq:5.13}
\eea
Then the multiple-Pomeron splitting expansion for nuclear diffractive amplitude 
takes the form
%%%%%%%%%%%%%%%%%%%%%%%%%%%%%         
\begin{eqnarray}        
\Phi_0^{(A)}(z,\bk,\bDelta)     
= 2  m_{f}\, \int d^2\bb e^{\displaystyle -i\bb \bDelta}          
 \left\{ \Psi_\pi(z,\bk)          
\left[ 1 - \exp \left( -{\sigma_{eff}(\bk^2)\over 2}  T_A(\bb) \right)\right]          
\right. \nonumber \\          
- \left. \sum_{j\geq 1}           
   \int d^2\bkappa \Psi_{\pi}(z,\bkappa)          
\phi_{WW}(\bk - \bkappa)\right\}\, .         
\label{eq:5.14}         
\end{eqnarray} 
A comparison with (\ref{eq:5.2}) suggests that         
$\phi_{WW}(\bkappa)$ can be identified with the unintegrated 
nuclear Weizs\"acker-Williams 
glue per unit area
in the impact parameter plane \cite{Saturation}. In (\ref{eq:5.11})
we kept an explicit dependence on the optical thickness of the
nucleus.

%-------------------- Section 6

\section{Nuclear dilution and broadening of the unintegrated 
Weizs\"acker-Williams glue of nuclei}

On the one hand, according to \cite{NSSdijet,Saturation} the multiple convolutions
$f^{(j)}(\bkappa )$ have a meaning of the collective unintegrated
gluon structure function of $j$ nucleons at the same impact parameter
the Weizs\"acker-Williams gluon fields of which overlap spatially
in the Lorentz contracted nucleus. On the other hand, these convolutions of can also
be viewed as a 
random walk in which $f(\bkappa )$ describes the single
walk distribution. 
To the lowest order in pQCD, the large $\bkappa ^2$ behavior is
\beq
f(\bkappa ) \propto {\alpha_S(\bkappa ^2)\over (\bkappa^2)^2}
\label{eq:6.1}
\eeq
The QCD evolution effects enhance $f(\bkappa )$ at large $\bkappa ^2$,
the smaller is $x$ the stronger is an enhancement. 
Because $f(\bkappa)$ decreases very slowly, we encounter
a manifestly non-Gaussian random walk. For instance, 
as was argued in \cite{NSSdijet}, a $j$-fold walk to large 
$\bkappa^2$ is realized by one large walk, $\bkappa_1^2 \sim \bkappa^2$, 
accompanied by $(j-1)$ small walks. We simply cite here the
main result \cite{NSSdijet}
\beq
f^{(j)}(\bkappa )=j\cdot f(\bkappa )\left[1+ {4\pi^2(j-1)\gamma^2
\over 
N_c\sigma_0\bkappa^2}
\cdot G(\bkappa^2)\right]
\label{eq:6.2}
\eeq
where $G(\bkappa^2)$ is the conventional integrated gluon structure function
and 
$\gamma$ is an exponent of the large-$\bkappa^2$ asymptotics
of the differential glue in the proton, 
\beq
f(\bkappa ) \propto {1\over (\bkappa^2)^{\gamma}}
\label{eq:6.3}
\eeq
Then the hard tail of unintegrated nuclear glue per bound nucleon 
can be calculated parameter free:
\bea
f_{WW}(\bkappa ) 
= f(\bkappa )\left[1+ {2 C_A\pi^2\gamma^2\alpha_S(r)T(\bb)\over 
C_F N_c \bkappa^2}
G(\bkappa^2)\right] 
\label{eq:6.4}
\eea
In the hard regime the differential nuclear glue is not
shadowed, furthermore, because of the manifestly positive-valued
and model-independent nuclear higher twist correction it 
exhibits nuclear antishadowing property \cite{NSSdijet}.

Now we present the arguments in favor of the soft-$\bkappa$ behavior
\beq
f^{(j)}(\bkappa) \approx {1\over \pi} {Q_j^2 \over (\bkappa^2 + Q_j^2 )^2}
={1\over Q_j^2} \chi({\bkappa^2 \over Q_j^2})
\label{eq:6.5}
\eeq
with the width 
\beq
Q_j^2 \sim  jQ_{0}^2
\label{eq:6.6}
\eeq
In an evolution of $f^{(j)}(\bkappa )$ with $j$ at moderate $\bkappa ^2$,
\beq
f^{(j+1)}(\bkappa ^2) =\int d^2\bk f(\bk^2)f^{(j)}((\bkappa -\bk)^2)
\label{beq:6.7}
\eeq
the function $f(\bk^2)$ is a steep one compared to a smooth and 
broad $f^{(j)}((\bkappa -\bk)^2)$, so that we can expand
\bea
&&f^{(j)}((\bkappa -\bk)^2)=\nonumber\\
&&f^{(j)}(\bkappa ^2) + 
{df^{(j)}(\bkappa ^2) \over d\bkappa^2}[\bk^2 -2\bkappa \bk]
+{1\over 2}{d^2f^{(j)}(\bkappa^2) \over( d\bkappa^2)^2}4(\bkappa \bk)^2
\nonumber\\
\Longrightarrow &&f^{(j)}(\bkappa^2) + 
\left[{df^{(j)}(\bkappa ^2) \over d\bkappa^2}
+\bkappa^2 {d^2f^{(j)}(\bkappa ^2) \over( d\bkappa^2)^2}\right]\bk^2 \nonumber\\
=&& f^{(j)}(\bkappa ^2) + 
{d \over d\bkappa^2}\left[\bkappa^2{df^{(j)}(\bkappa ^2) \over d\bkappa^2}
\right]\bk^2
\label{eq:6.8}
\eea
Here $\Longrightarrow$ indicates an azimuthal averaging. Our expansion 
was valid for $\bk^2 \lsim Q_j^2$, hence 
\bea
f^{(j+1)}(\bkappa^2)=f^{(j)}(\bkappa^2) + 
{d \over d\bkappa^2}\left[\bkappa^2{df^{(j)}(\bkappa ^2) \over d\bkappa^2}
\right]
\int^{Q_j^2}d^2\bk f(\bk^2)\bk^2 \nonumber\\
=
f^{(j)}(\bkappa^2) + g(j){d \over d\bkappa^2}
\left[\bkappa^2{df^{(j)}(\bkappa ^2) \over d\bkappa^2}
\right]
\label{eq:6.9}
\eea
where 
\beq
g(j) = \int^{Q_j^2}d^2\bk \bk^2f(\bk^2) = 
{ 4\pi^2 \over N_c\sigma_{0}}G(Q_j^2)
\label{eq:6.10}
\eeq
is a smooth function of $j$. 

For small $\bkappa^2$ the recurrence relation amounts to
\bea
{ Q_{j+1}^2 - Q_j^2  \over  Q_{j+1}^2 Q_j^2} = -{1\over Q_j^4}{\chi'(0) \over \chi(0)}g(j)
\label{eq:6.11}
\eea
which, for large $j$, reduces to a differential equation
\beq
{d Q_{j}^2 \over dj}  = -{\chi'(0)\over \chi(0)} g(j)
\label{eq:6.12}
\eeq
with the solution
\beq
Q_{j}^2 = - {\chi'(0))\over \chi(0)}
\int^{j} dj'g(j') \approx  - jg(j) {\chi'(0)\over \chi(0)} \approx jQ_0^2.
\label{eq:6.13}
\eeq
The expansion (\ref{eq:6.8}) holds up to the terms $\propto \bkappa^2$
and its differentiation at $\bkappa^2=0$ gives a similar constraint 
on the $j$-dependence of $Q_{j}^2$. The soft
parameters $Q_0^2$ and $\sigma_0 $ are related to the integrated
glue of the proton in the soft region,
\beq
Q_{0}^2\sigma_0 \sim {2\pi^2 \over N_c} G_{soft}\,, ~~~G_{soft}\sim 1\,.
\label{eq:6.14}
\eeq
We conclude by the observation that when extended to $\bkappa^2 \gsim 
Q_{j}^2$, the parameterization (\ref{eq:6.5}), (\ref{eq:6.6}) has the 
behavior $jQ_{0}^2/(\bkappa^2)^2$ which nicely matches the $j$-dependence 
of the leading twist term in the hard asymptotics (\ref{eq:6.4}).

An approximation (\ref{eq:6.5}), (\ref{eq:6.6}) is corroborated by
the numerical results for the $j$-dependence of convolutions 
calculated for the unintegrated gluon structure function of the proton
from ref. \cite{INDiffGlue}, see fig.~6.

%-------------------   figure 6

\begin{figure}[!htb]
\begin{center}
%\vspace{1.5cm}\\
   \epsfig{file=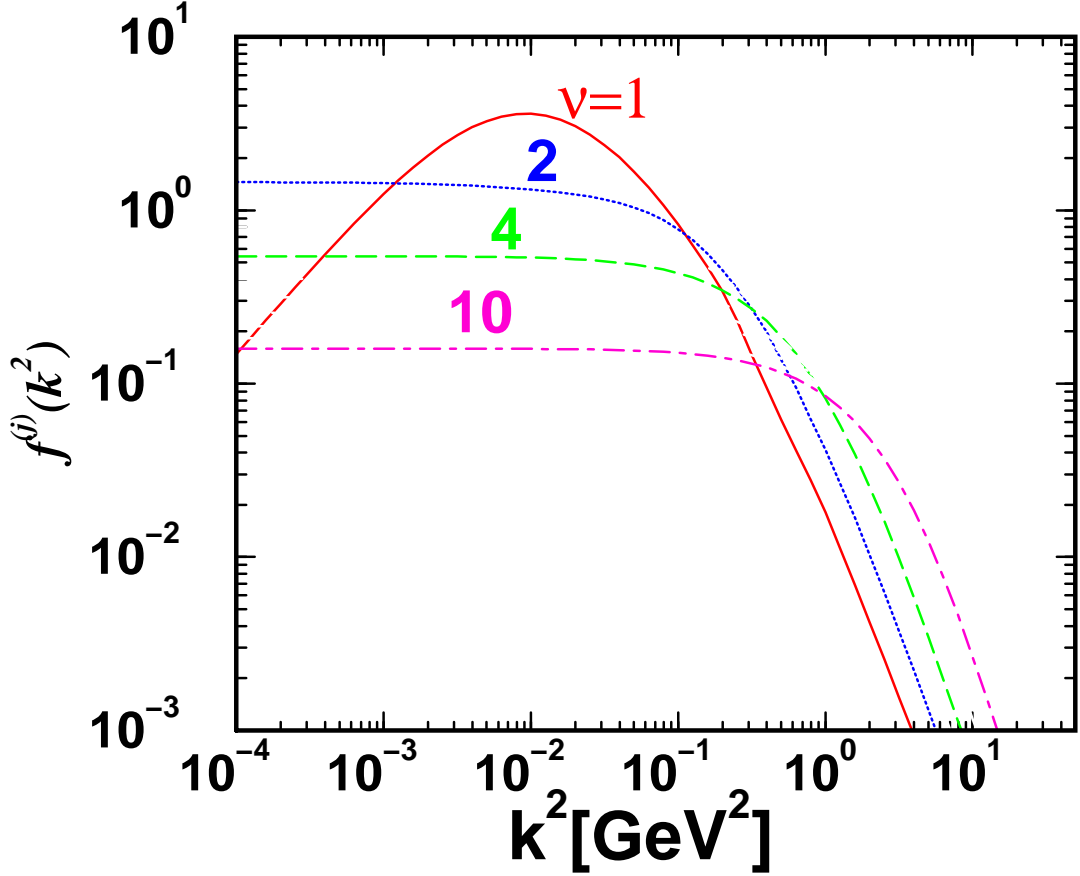,width=14cm}
\caption{\it The dilution for soft momenta and broadening for hard
momenta of the multiple convolutions $f^{(j)}(\bkappa)$ .}
\end{center}
\end{figure}

For heavy nuclei the dominant contribution to $\phi_{WW}(\bkappa)$ comes
from $j\approx \nu_A(\bb)$, and the gross features of the
WW nuclear glue in the soft region are well reproduced by 
\bea
\phi_{WW}(\bkappa) \approx  {1\over \pi}  {Q_A^2 \over (\bkappa^2
+Q_{A}^2)^2}\, , 
\label{eq:6.15} \eea 
where the saturation scale $Q_A^2 =  \nu_A(\bb)  Q_0^2 \propto A^{1/3}\, . $ 

Notice a strong  nuclear dilution of soft WW glue, 
\beq
\phi_{WW}(\bkappa) \propto 1/Q_A^2 \propto
A^{-1/3}
\label{eq:6.16}
\eeq
which must be contrasted to the $A$-dependence of hard glue, cf. (\ref{eq:6.2}),
(\ref{eq:6.4}), 
\beq
\phi_{WW}(\bkappa) = \nu_A(\bb) f_{WW}(\bkappa) \propto A^{1/3}\times
(1 + A^{1/3}\times(HT)+...)\, .
\label{eq:6.17}
\eeq

Take the platinum target, $A=192$. The numerical estimates show that
for $(q\bar{q})$ color dipoles in the average DIS  on this target
$
\nu_{A} \approx 4
$ and 
$
Q_{3A}^2  \approx 0.8 {\rm GeV}^2\, .
$
For the $q\bar{q}g$ Fock states of the photon, which behave predominantly 
like  the dipole made of the two octet color charges, the saturation scale
is larger by the factor $C_A/C_F = 9/4$,
\beq
Q_{8A}^2 = {C_A \over C_F}Q_{3A}^2\, .
\label{eq:6.18}
\eeq
The numerical estimates for the the average DIS on the Pt 
target give $ Q_{8A}^2 \approx 2.2$ GeV$^2$.

%--------------------------- Section 7

\section{Back to the hard diffraction of pions: the interpretation 
of the E791 data}

The detailed numerical analysis for the kinematics of the E791 experiment         
is found in \cite{NSSdijet}. In fig.~7 we only show the numerical result          
of the $\bk$ dependence of the dijet cross section with the          
(unnormalized)  data from E791. We note in passing that the          
region of jet momenta $k \lsim 1.5$ GeV is contaminated by          
diffractive excitation of heavy mesons $a_1,\pi'$, etc, and in this          
region the use of plane wave parton model formulas is not         
warranted. Our calculations for smaller $k$ only serve to give an     
idea on how the pion splitting dominates at small $k$ and how the     
pomeron splitting mechanism takes over beyond the dip in $\Phi_{0}^2$     
and $\bPhi_{1}^2$. The latter is a higher twist term.
In the pomeron splitting dominance      
region of $k > 1.5$ GeV we find good agreement         
with the experimentally observed $\bk$-dependence. 
We wish to warn the reader however, that for the kinematics     
of E791 the dijet cross section receives a huge contribution         
from the antishadowing nuclear higher twist correction \cite{NSSdijet}.          

%-------------------   figure 7

\begin{figure}[!htb]         
\begin{center}         
\epsfig{file=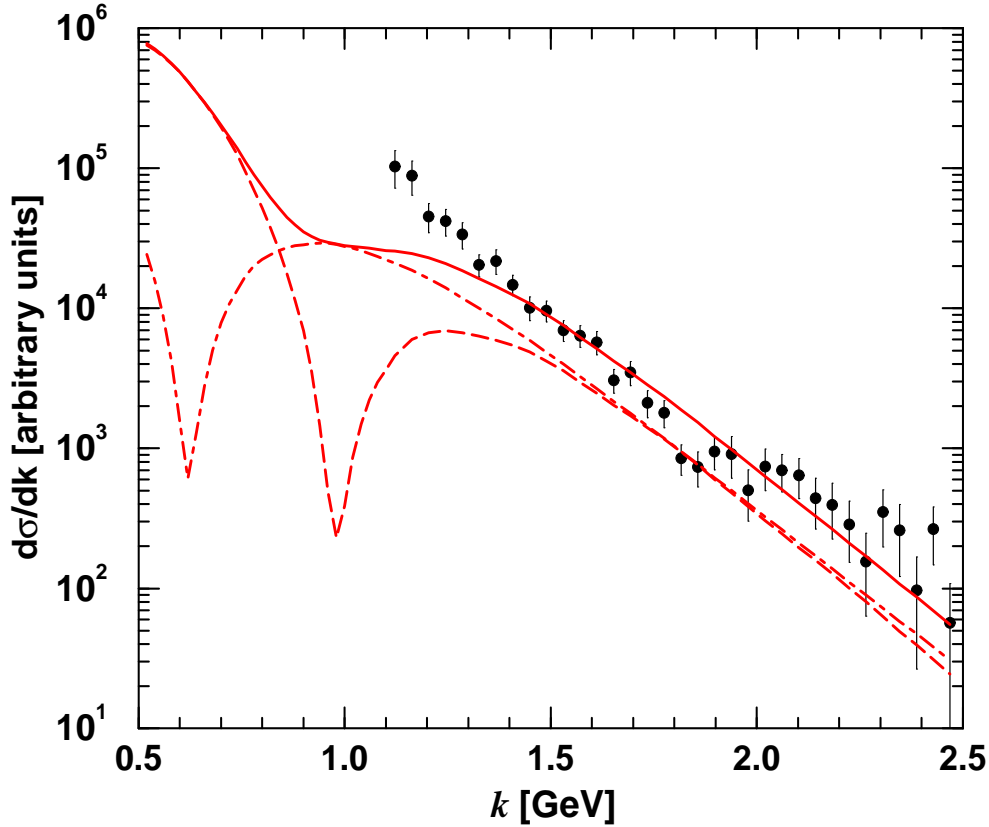, height = 13.0cm, width=11.0cm,angle=270}         
\end {center}         
\caption{\it The E791 data \protect\cite{Ashery} for the differential diffractive         
dijet cross section $d\sigma/dk$ for the $^{196}$Pt target with         
the theoretical calculations. The data are not normalized.         
The dash--dotted line shows the contribution of the helicity         
amplitude $\Phi^{(A)}_0$, the dashed line is         
the contribution from $\bPhi^{(A)}_1$. The solid line is the          
total result.}         
\label{fig7}         
\end{figure}

The effect of antishadowing nuclear higher twist correction is especially
obvious in fig.~8, where we show the exponent $\alpha$ of the $A$-dependence 
$\sigma \sim A^{\alpha}$. In the impulse approximation
\beq
\sigma \propto {A^2 \over R_A^2} \sim A^{4/3}
\label{eq:7.1}
\eeq
If the exponent $\alpha$ is defined from a comparison of the cross sections
for the carbon and platinum targets, then the impulse approximation gives
$\alpha \approx 1.44$. For hard dijets our antishadowing effect predicts
 a substantial excess of $\alpha$ over the impulse approximation value. 
 
\begin{figure}[!htb]         
\begin{center}         
\epsfig{file=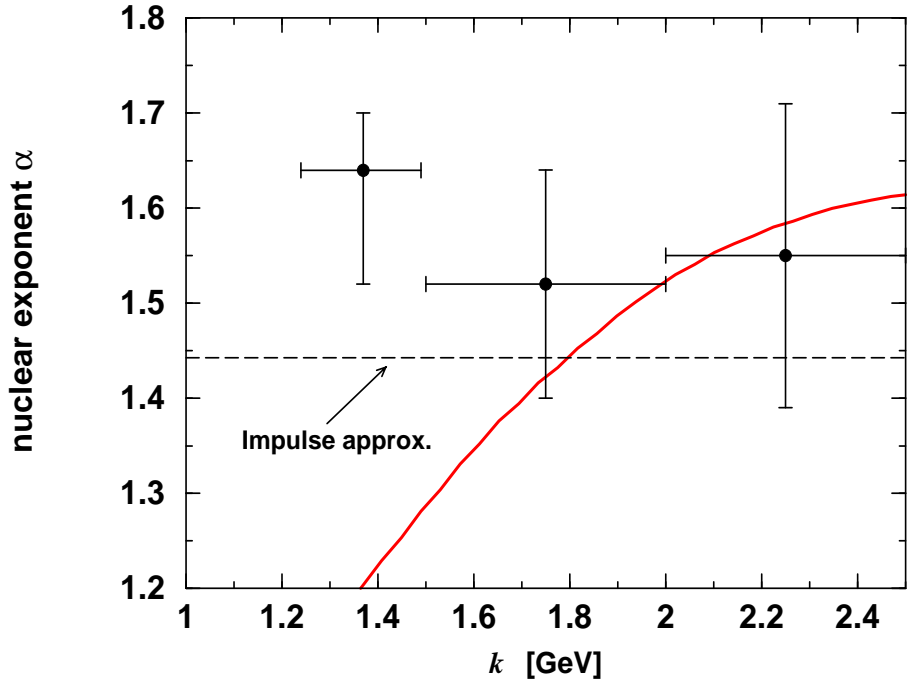, height = 12.0cm, width=9.0cm,angle=270}
\caption{\it The E791 data \protect\cite{Ashery} for the exponent of
the $A$ dependence of the jet cross section vs.          
the theoretical calculations shown by thge solid curve. } 
\end {center}                  
\label{fig2}         
\end{figure}

The NSS dominance
of the pomeron-splitting contribution for hard dijets
has fully been confirmed by the collinear factorization
NLO order analysis of Chernyak et al. \cite{Chernyak}
and Braun et al. \cite{Regensburg}. The NLO correction to the
NSS amplitude is found to be proportional to the asymptotic
distribution amplitude and  numerically quite substantial, so
that the experimental data by E791 \cite{Ashery} can not
distinguish between the asymptotic and double-humped distribution
amplitudes. According to NSS \cite{NSSdijet} realistic model
distributions do not differ much from the asymptotic one, though. To
our view, the
only caveat in the interpretation of the NLO results is that
the issue of partial reabsorption
of these corrections into the evolution/renormalization of the
pion distribution amplitude still needs further scrutiny. 
Anyway, there emerges a consistent pattern of diffraction of pions
into hard dijets and in view of these findings the claims by Frankfurt
et al. \cite{FMS} that the diffractive amplitude is proportional
to the integrated gluon structure function of the target must be
regarded null and void. (Hopefully, some day the E791 collaboration
shall report on the interpretation of their results within the
correct formalism.)

%--------------------  Section 8

\section{Nuclear partons in the saturation regime}

Here we apply the results of sections 4 and 5 to the calculation
of FS single-parton (jet) spectrum for nuclear targets.
On the one hand, making use of the NSS representation, the total
nuclear photoabsorption cross section 
(\ref{eq:5.8}) can be cast in the form 
\beq
\sigma_{A}  = \int d^2\bb\int dz \int {d^2\bp\over (2\pi)^2} \int
d^2\bkappa\phi_{WW}(\bkappa) \left|\langle \gamma^* |\bp\rangle -
\langle \gamma^* |\bp-\bkappa\rangle \right|^2 
\label{eq:8.1} 
\eeq
which has a profound  semblance to (\ref{eq:2.5}) and one is tempted
to take the differential form of (\ref{eq:8.1}) as a definition of
the IS sea quark density in a nucleus (hereafter we consider one
flavour and take $e_f^2$ the factor out of the photon's wave function
and cross sections): 
\bea {d\bar{q}_{IS} \over
d^2\bb d^2\bp} = {1\over 2}\cdot{Q^2 \over 4\pi^2 \alpha_{em}}
\cdot{d\sigma_A \over d^2\bb d^2\bp}\, . 
\label{eq:8.2} 
\eea 
In
terms of the WW nuclear glue, all intranuclear multiple-scattering
diagrams of fig.~3g sum up to precisely the same four diagrams
fig.~3a-3d as in DIS off free nucleons. Furthermore, one can argue
that the small-$x$ evolution of the so-defined IS nuclear sea is
similar to that for a free nucleon sea. Although $\bp$ emerges
here just as a formal Fourier parameter, we shall demonstrate that
it can be identified with the momentum of the observed final state
antiquark.

On the other hand, making use of the NSS representation, after
some algebra one finds 
\bea 
&&\exp[-{1\over
2}\sigma(\br-\br')T(\bb)]- \exp[-{1\over
2}[\sigma(\br)+\sigma(\br')]T(\bb)] = \nonumber\\
&&\int d^2\bkappa
\phi_{WW}(\bkappa)
 \{(\exp[i\bkappa(\br-\br')]-1) +
(1-\exp[i\bkappa\br])+(1-\exp[i\bkappa\br'])\}\nonumber\\
&& - \int
d^2\bkappa \phi_{WW}(\bkappa)(1-\exp[i\bkappa\br]) \int
d^2\bkappa' \phi_{WW}(\bkappa')(1-\exp[i\bkappa'\br'])~~~
\label{eq:8.3} 
\eea
so that the single-quark spectrum (\ref{eq:4.12} takes the form 
\bea 
{d \sigma_{in}\over d^2\bb d^2\bp dz } &=&
{1 \over (2\pi)^2}\left\{ \int  d^2\bkappa
\phi_{WW}(\bkappa)\left|\langle \gamma^* |\bp\rangle -
\langle \gamma^* |\bp-\bkappa\rangle \right|^2 \right. \nonumber\\
&& - \left. \left|\int d^2\bkappa\phi_{WW}(\bkappa) (\langle
\gamma^* |\bp\rangle - \langle \gamma^*
|\bp-\bkappa\rangle )\right|^2\right\} 
\label{eq:8.4} \\
%\eea
%\bea
{d \sigma_{D}\over d^2\bb d^2\bp dz }   &=&  {1 \over (2\pi)^2}
\left|\int d^2\bkappa\phi_{WW}(\bkappa) (\langle \gamma^*
|\bp\rangle - \langle \gamma^* |\bp-\bkappa\rangle)\right|^2 \, .
\label{eq:8.5} 
\eea 
Putting the inelastic and
diffractive components of the FS quark spectrum together, we
evidently find the FS parton density which exactly coincides with
the IS parton density (\ref{eq:8.2}) such that $\bp$ is indeed the
transverse momentum of the FS sea quark. The interpretation of
this finding is not trivial, though.

Consider first the domain of $\bp^2 \lsim Q^2 \lsim Q_A^2$ such
that the nucleus is opaque for all color dipoles in the photon.
Hereafter we assume that the saturation scale $Q_A^2$ is so large
that $\bp^2,Q^2$ are in the pQCD domain and neglect the quark
masses. In this regime the nuclear counterparts of the crossing
diagrams of figs. 3b,3d,3f  can be neglected. Then, in the
classification of \cite{NSSdijet}, diffraction will be dominated by the
contribution from the Landau-Pomeranchuk diagram of fig.~3e with
the result 
\bea 
\left. {d\bar{q}_{FS} \over d^2\bb
d^2\bp}\right|_D &=& {1\over 2}\cdot{Q^2 \over 4\pi^2 \alpha_{em}}
\cdot{d\sigma_D \over d^2\bb d^2\bp} \nonumber\\
&\approx& {1\over 2}\cdot{Q^2 \over 4\pi^2 \alpha_{em}} \cdot\int
dz \left| \int d^2\bkappa\phi_{WW}(\bkappa) \right|^2
\left|\langle \gamma^* |\bp\rangle\right|^2 \nonumber\\
&\approx& {N_c \over
4\pi^4}\, . 
\label{eq:8.6} 
\eea 
Up to now we specified neither the
wave function of the photon nor the spin  nor the color
representation of charged partons, only the last result in
(\ref{eq:8.6}) makes an explicit use of the conventional spin-${1\over
2}$ partons.
Remarkably, diffractive DIS measures the momentum distribution of
quarks and antiquarks in the $q\bar{q}$ Fock state of the photon.
We emphasize that this result, typical of the Landau-Pomeranchuk
mechanism, is a completely generic one and would hold for any beam
particle such that its coupling to hard colored partons is weak. In
contrast to diffraction off free nucleons
\cite{NZ92,NZsplit,GNZcharm}, diffraction off opaque nuclei is
dominated by the anti-collinear splitting of hard gluons into soft
sea quarks, $\bkappa^2 \gg \bp^2$. Precisely for this reason one
finds the saturated FS quark density, because the nuclear dilution
of the WW glue is compensated for by the expanding plateau. The
result (\ref{eq:8.6}) has no counterpart in DIS off free nucleons
because diffractive DIS off free nucleons is negligibly small even
at HERA, $\eta_D \lsim $ 6-10 \% \cite{GNZ95}.

The related analysis of the FS quark density for truly inelastic
DIS in the  same domain of $\bp^2 \lsim Q^2 \lsim Q_A^2$ gives
\bea \left.{d\bar{q}_{FS} \over d^2\bb d^2\bp}\right|_{in} &=&
{1\over 2}\cdot{Q^2 \over 4\pi^2 \alpha_{em}} \cdot\int dz \int
d^2\bkappa \phi_{WW}(\bkappa)
\left|\langle \gamma^* |\bp-\bkappa\rangle \right|^2 \nonumber\\
 &=&
{Q^2 \over 8\pi^2 \alpha_{em}}\phi_{WW}(0)
\int^{Q^2} d^2\bkappa \int dz \left|\langle \gamma^* |\bkappa\rangle \right|^2
\nonumber\\
&=& {N_c \over 4\pi^4}\cdot {Q^2 \over Q_A^2}\cdot\theta(Q_A^2-\bp^2)\, .
\label{eq:8.7}
\eea
It describes final states with color excitation of a nucleus,
but as a function of the photon wave
function and nuclear WW gluon distribution it is completely
different from (\ref{eq:2.5}) for free nucleons. The $\theta$-function simply
indicates that the plateau for inelastic DIS extends up to $\bp^2
\lsim Q^2_A$.
For $Q^2 \ll Q_A^2$ the inelastic plateau
contributes little to the transverse momentum distribution of
soft quarks, $\bp^2 \lsim Q^2$, but the inelastic plateau extends way beyond
$Q^2$ and its integral contribution to the spectrum of FS
quarks is exactly equal to that from diffractive DIS. Such a  two-plateau
structure of the FS quark spectrum is a new finding and has not been
considered before.

\begin{figure}[!htb]
   \centering
   \epsfig{file=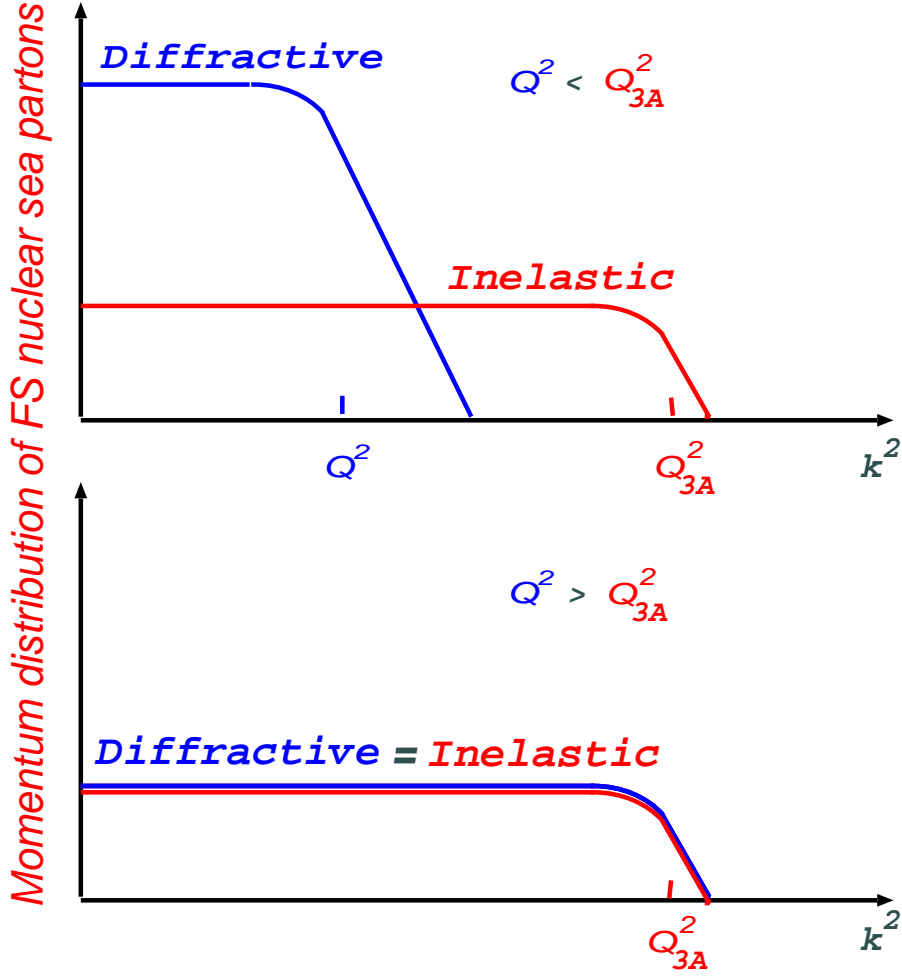,width=12cm}
\caption{\it The two-plateau structure of the 
momentum distribution of FS quarks: (i)
$Q^2 \lsim Q_{3A}^2$ : The inelastic  plateau is much broader than
the diffractive one. (ii)
$Q^2 \gsim Q_{3A}^2$ : inelastic  plateau is identical to diffractive one.
} 
\end{figure}

Now notice, that  in the opacity regime the diffractive FS parton
density coincides with the contribution $\propto |\langle
\gamma^*|\bp\rangle|^2$ to the IS sea parton density from the
spectator diagram of fig. 3a, whereas the FS parton density for truly
inelastic DIS coincides with the contribution to IS sea partons
from the diagram of fig.~3c. In this regime the contribution from the crossing
diagrams of figs. 3b,d is negigibly small.

Our results (\ref{eq:8.6}) and (\ref{eq:8.7}), especially nuclear broadening
and unusually strong $Q^2$ dependence of the FS/IS parton density from
truly inelastic DIS, demonstrate clearly a distinction between diffractive
and inelastic DIS. Our considerations can readily be extended to the
spectrum of soft quarks, $\bp^2 \lsim Q_A^2$, in hard photons, $Q^2 \gsim Q_{A}^2$.
In this case the result (\ref{eq:8.6}) for diffractive DIS is retained,
whereas in the numerator of the result (\ref{eq:8.7}) for truly inelastic
DIS one must substitute $Q^2 \to Q_{A}^2$, so that in this case
$dq_{FS}|_{D} \approx dq_{FS}|_{in}$ and $dq_{IS} \approx
2dq_{FS}|_{D}$. The evolution
of soft nuclear sea, $\bp^2 \lsim Q_{A}^2$, is entirely driven by
an anti-collinear splitting of the NSS-defined WW nuclear glue into the sea
partons.

The early discussion of the FS quark density in the saturation
regime is due to Mueller \cite{Mueller}. Mueller focused on $Q^2
\gg Q_A^2$ and discussed  neither a distinction between
diffractive and truly inelastic DIS nor a $Q^2$ dependence and
broadening (\ref{eq:8.7}) for truly inelastic DIS at $Q^2 \lsim
Q_A^2$.

 A comparison with the IS nuclear parton
densities which evolve from the NSS-defined WW nuclear glue shows
an exact equality of the FS and IS parton densities. The
plateau-like saturated nuclear quark density is suggestive of the
Fermi statistics, but our principal point that for any projectile
which interacts weakly with colored partons the saturated density
measures the momentum distribution in the $q\bar{q} , gg,...$ Fock
state of the projectile disproves the Fermi-statistics
interpretation. The spin and color multiplet of colored partons
the photon couples to is completely irrelevant, what only counts
is an opacity of heavy nuclei. The anti-collinear splitting of WW
nuclear glue into soft sea partons is a noteworthy feature of the
both diffractive DIS and IS sea parton distributions. 
The
emergence of a saturated density of IS sea partons from the
nuclear-diluted WW glue is due to the nuclear broadening of the
plateau (\ref{eq:8.7}). Because the predominance of diffraction is
a very special feature of DIS \cite{NZZdiffr}, one must be careful
with applying the IS parton densities to, for instance, nuclear
collisions, in which diffraction wouldn't be of any significance.

% -----------------------   Section 9

\section{Signals of saturation in exclusive diffractive DIS}

A flat $\bp^2$ distribution of forward $q,\bar{q}$ jets 
in truly inelastic DIS in the saturation regime, see eq.~(\ref{eq:8.7}),
 must be contrasted 
to the standard DGLAP spectrum,
\beq
{d\bar{q} \over d^2\bp}= {2\alpha_{S}(\bp^2)G(\bp^2)\over 3\pi \bp^2}\, , 
\eeq
for the free nucleon
target. 

In the diffractive DIS the signal of saturation is much more
dramatic: a flat $\bp^2$ distribution of forward $q,\bar{q}$ jets 
in diffractive DIS in the saturation regime, see eq.~(\ref{eq:8.6}),
must be contrasted to the spectrum
\beq
{d\sigma_D \over d^2\bp }\propto {G^2(\bp^2) \over (\bp^2)^2}
\eeq  
in diffractive DIS off the free nucleon
target \cite{NZ92,NZsplit,GNZcharm}. Recall that it is $\bp^2$ 
which serves as a hard scale for diffractive dijets unless 
$\bp^2$ is so large that the pomeron splitting mechanism takes
over \cite{NZsplit,GNZcharm}.

\begin{figure}[!htb]
   \centering
   \epsfig{file=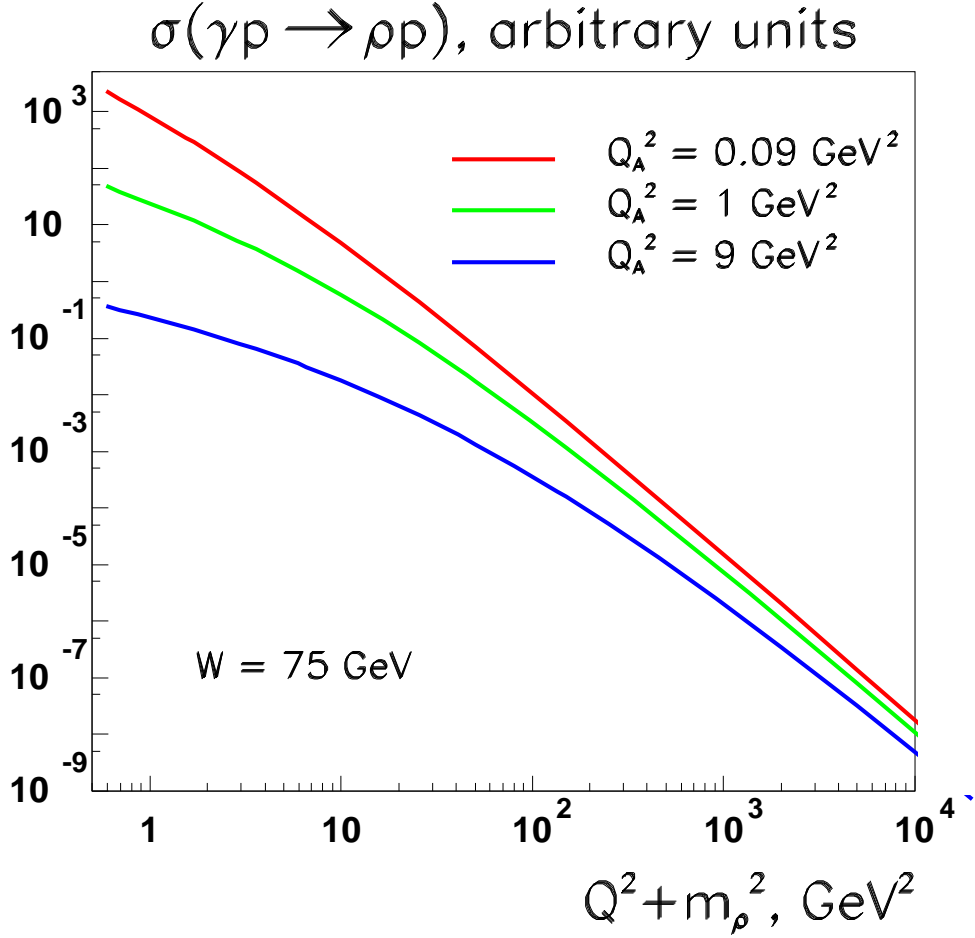,width=13.5cm}
\caption{ \it The modification of the $(Q^2+m_V^2)$ dependence 
of diffractive
 $\rho$ production from a free nucleon target (small saturation 
scale $Q_A^2$) to a nucleus (large $Q_A^2$) 
with the change of the saturation
scale $Q_A^2$.} 
\end{figure}

In the general case one must
compare the saturation scale $Q_A$ to the relevant hard scale
for the specific diffractive process. 
For instance, in the exclusive diffractive DIS, i.e., the vector meson
production, the hard scale for the proton target 
equals \cite{NNZscan}
\beq
\bar{Q}^2 \approx {1\over 4}(Q^2+m_V^2)
\label{eq:9.1}
\eeq
and the transverse cross section has been predicted to behave
as \cite{NNZscan}
\beq
\sigma_T \propto \left[{G(x,\bar{Q}^2)\over \bar{Q}^4}\right]^{2}
\label{eq:9.2}
\eeq
At $\bar{Q}^2 > Q_A^2$ the same would hold for nuclei too, 
but in the opposite case of $\bar{Q}^2 < Q_A^2$ the 
$\bar{Q}^2$-dependence is predicted to change to 
\beq
\sigma_T \propto \left[{G(x,\bar{Q}^2)\over Q_A^2\bar{Q}^2}\right]^{2}
\label{eq:9.3}
\eeq
The numerical results shown in Fig.~10 give an idea on how the
$(Q^2+m_V^2)$ dependence of exclusive diffraction into the
vector mesons changes from the free nucleon to nuclear target 
with the incresaing saturation scale.

As shown in \cite{GNZlong}, there is a duality correspondence 
between the  $(Q^2+m_V^2)$ dependence of 
exclusive production of vector mesons and the mass spectrum of
diffractive DIS into small-mass continuum states, i.e,
at large values, $\beta \to 1$, of the diffractive Bjorken
variable 
\beq
\beta = {Q^2 \over M^2 +Q^2}\, .
\eeq 
Specifically, for diffractive DIS off free nucleons the 
transverse diffractive structure function has the large-$\beta$ 
behavior
\cite{GNZcharm}
\beq
F_D \propto (1-\beta)^2\, ,
\eeq
whereas in diffractive DIS off nuclei in the saturation regime
we predict
\beq
F_D \propto (1-\beta)\, .
\eeq

% -------------------  Section 10

\section{Jet-jet (de)correlation in DIS off nuclear targets: hard jets}

The diagonalization of the $2\times 2$ matrix $\sigma_4$ derived
in section 3 is a
straightforward task, so that technically eqs. 
(\ref{eq:3.1}) and (\ref{eq:3.13})-(\ref{eq:3.18}) allow a
direct calculation of the jet-jet inclusive cross section
in terms of the color dipole cross section $\sigma(\br)$.
The practical evaluation of the 6-fold Fourier transform
is not a trivial task, though, and anlytical evaluations
are called upon. Here we focus on production of forward hard 
jets with the momenta $\bp_{\pm}^2 \gsim Q_A^2$, where $Q_A$ 
is the saturation scale. Such a hard jets are produced from 
interactions with the target nucleus of small color dipoles 
in the incident photon such that diffractive nuclear attenuation 
effects can be neglected. In this case useful and intuitively 
transparent analytic results for the jet-jet (de)correlation 
can be obtained.

Before proceeding to this approximation, it is convenient to 
introduce the average impact parameter
\beq
\bb = {1\over 4}( \bb_+ + \bb_+' + \bb_- + \bb_-')\, ,
\label{eq:10.1}
\eeq
(one should nor confuse $\bb$ with the center of gravity of
color dipoles where the impact parameters $\bb_{\pm}$ and $\bb_{\pm}'$ 
must be weighted with $z_{\pm}$) and 
\beq
\bs = \bb_+ - \bb_+' \, ,
\label{eq:10.2}
\eeq
for the variable conjugate to the decorrelation momentum,   
in terms of which
\bea
\bb_+ - \bb_-' = \bs + \br'\, ,
\label{eq:10.3}
\eea
\bea
\bb_- - \bb_+' = \bs - \br \, ,
\label{eq:10.4}
\eea
\bea
\bb_- - \bb_-' = \bs -\br + \br'\, .
\label{eq:10.5}
\eea
We also reproduce here the matrix elements of $\sigma_4$ in
this basis:
\bea
\langle 11|\sigma_4|11\rangle = \sigma(\br)+\sigma(\br')
\label{eq:10.6} 
\eea
\bea
\langle 11|\sigma_4|88\rangle = 
{1\over \sqrt{N_c^2-1}}
\left[\sigma(\bs)-
\sigma(\bs + \br')- \sigma(\bs -\br) + \sigma(\bs - \br + \br')\right]
\label{eq:10.7} 
\eea
\bea
\langle 88|\sigma_4|88\rangle & =& 
{N_c^2 -2 \over N_c^2-1}[
[\sigma(\bs)+
\sigma(\bs -\br + \br')]\nonumber\\
&+&{2 \over N_c^2-1}[\sigma(\bs + \br')+ 
\sigma(\bs - \br)] \nonumber\\ 
&-&{1\over N_c^2-1}
[\sigma(\br)+\sigma(\br')]
\label{eq:10.8} 
\eea

Hard jets correspond to $|\br|, |\br'| \ll |\bs|$.
Then the two eigenvalues are 
\bea
\Sigma_2=\langle 11|\sigma_4|11\rangle \approx 0
\label{eq:10.9} 
\eea
and 
\bea
\Sigma_1=\langle 88|\sigma_4|88\rangle \approx 
2{N_c^2 \over N_c^2-1}\sigma(\bs) = 2\lambda_c \sigma(\bs)
\label{eq:10.10} 
\eea
where $\lambda_c = N_c^2/(N_c^2-1)$. Because of (\ref{eq:10.9})
only the second term,
 $\propto \langle 11| \sigma_4 | 88\rangle$, must be kept in the
Sylvester expansion (\ref{eq:3.19}).
It is convenient to define the nuclear distortion factor
\bea
D_A(\bs,\br,\br') =  { 2 \over (\Sigma_2 - \Sigma_1)T(\bb)}
\left\{\exp\left[-{1\over 2}\Sigma_1 T(\bb)\right]-
\exp\left[-{1\over 2}\Sigma_2 T(\bb)\right]\right\}
\label{eq:10.11}
\eea
such that to the leading order in nuclear thickness, i.e.,
in the impulse approximation, $D_A(\bs,\br,\br')=1$. Then, the impulse
approximation cross section per unit area in the impact parameter
plane takes the form (we expose the relevant steps in some detail) 
\bea
&&{d\sigma_{IA} \over d^2\bb dz d^2\bp_+ d^2\bp_-} =
{-1\over 2(2\pi)^4} \int d^2\bs d^2\br d^2\br' \nonumber\\
&&\times \exp[-i(\bp_+ +\bp_-)\bs
+i\bp_-(\br' -\br)] \Psi^*(\br') 
\Psi(\br) \nonumber\\
&& \times T(\bb)\sqrt{N_c^2 -1} \langle 11| \sigma_4 | 88\rangle D_A(\bs,\br,\br',\bb)
\nonumber\\
&&=
{1\over 2 (2\pi)^4}T(\bb) \int d^2\bs d^2\br d^2\br'
\exp[-i(\bp_+ +\bp_-)\bs
+\bp_-(\br'-\br)]
\nonumber\\
&& \times \Psi^*(\br')\Psi(\br)
\left[
\sigma(\bs)-
\sigma(\bs + \br')- \sigma(\bs -\br) + \sigma(\bs - \br + \br') \right]
\label{eq:10.12}
\eea
Here we employ the integral representation (\ref{eq:2.1}):
\bea
\left[
\sigma(\bs)-
\sigma(\bs + \br')- \sigma(\bs -\br) + \sigma(\bs - \br + \br') \right]
\nonumber\\
=
\alpha_S \sigma_{0} \int d^2\bDelta f(\bDelta)
\exp[i\bDelta\bs]
\left\{1 - \exp[i\bDelta \br']\right\}
\left\{1 - \exp[-i\bDelta \br]\right\}\, ,
\label{eq:10.13}
\eea
which leads to the Impulse Approximation result
\bea
&&{d\sigma_{IA} \over d^2\bb dz d^2\bp_+ d^2\bp_-} 
={1\over 2 (2\pi)^4}\alpha_S \sigma_{0}T(\bb) \int d^2\bs d^2\br d^2\br' 
d^2\bDelta f(\bDelta)\nonumber\\
&&\times \exp[i(\bDelta -\bp_+ -\bp_-)\bs]
\exp[
i\bp_-(\br'-\br)] \nonumber\\
&&\times \Psi^*(\br')\left\{1 - \exp[i\bDelta \br']\right\}\Psi(\br)
\left\{1 - \exp[-i\bDelta \br]\right\}\, .
\label{eq:10.14}
\eea
The $d^2\bs$ integration entailes $\bDelta = \bp_+ -\bp_-$, the 
remaining Fourier transforms are straightforward and give precisely 
(\ref{eq:2.9}) times $T(\bb)$: 
\bea
{d\sigma_{IA} \over d^2\bb dz d^2\bp_+ d^2\bDelta}= T(\bb) 
{d\sigma_N \over d^2\bb dz d^2\bp_+ d^2\bDelta}
\label{eq:10.15}
\eea

 The nuclear distortion factor
takes a simple form
\bea
D_A(\bs,\bb)={\exp \left[-{1\over 2}\Sigma_1 T(\bb)\right]-1
\over {1\over 2}\Sigma_1 T(\bb)] }=  
\int_0^1 d \beta \exp\left[-{1\over 2}\beta \Sigma_1 T(\bb)\right] \nonumber\\
= \int_0^1 d \beta  \int d^2\bkappa 
\Phi(2\beta \lambda_c  \nu_A(\bb),\bkappa)\exp(i\bkappa \bs)
\label{eq:10.16}
\eea
The introduction of this distortion factor into (\ref{eq:10.14}) is straightforward
and gives our central result for the hard jet-jet inclusive cross section:
\bea
{d\sigma_{in} \over d^2\bb dz d^2\bp_+ d^2\bDelta}= T(\bb)
\int_0^1 d \beta \int d^2\bkappa \Phi(2\beta \lambda_c  \nu_A(\bb),\bDelta - \bkappa)
{d\sigma_{N} \over dz d^2\bp_+ d^2\bkappa }\, .
\label{eq:10.17} 
\eea  
It has a probabilistic form of a convolution of the differential cross 
section on a free nucleon target with $\Phi(2\beta \lambda_c  \nu_A(\bb),\bkappa)$. 
Here $\beta$ has a meaning of the fraction of the nuclear thickness 
which the $(q\bar{q})$
pair propagates in the color octet state.

In the practical calculations one can use an explicit expansion
\bea
\int_0^1 d \beta \Phi(2\beta \lambda_c  \nu_A(\bb),\bDelta) = 
\sum_{j=0}^{\infty} {1 \over j!} {\gamma(j+1,2\beta \lambda_c  \nu_A(\bb)) \over 
2\beta \lambda_c  \nu_A(\bb)} 
f^{(j)}(\bDelta)\, ,
\label{eq:10.18}
\eea
where $\gamma(j,x)$ is an incomplete Gamma-function.
For heavy nuclei the dominant contribution comes from $j\sim \nu_{A}$. 
Neglecting the small departure of $\lambda_c$ from unity, for a heavy nucleus
we can approximate 
\bea
&&\int_0^1 d \beta \Phi(2\beta \lambda_c  \nu_A(\bb),\bDelta) \approx
\Phi( \lambda_c  \nu_A(\bb),\bDelta) \nonumber\\
&&\approx \Phi(\nu_A(\bb),\bDelta)\approx
{1\over \pi} {Q_A^2 \over (\bDelta^2 + Q_A^2 )^2}
\label{eq:10.19}
\eea
and evaluate the gross features of a jet-jet decorrelation analytically.
First, notice that $f(\bkappa)$ which enters the free nucleon cross section
(\ref{eq:2.9}) is a steep function of $\bkappa$ compared to a broad
distribution (\ref{eq:10.19}). Then,  
\bea
&&{d\sigma_{in} \over d^2\bb dz d^2\bp_+ d^2\bDelta} \approx T(\bb) {Q_A^2 \over 
\pi (\bDelta^2 + Q_A^2 )^2} \int d^2\bkappa 
{d\sigma_{N} \over dz d^2\bp_+ d^2\bkappa } \nonumber \\
&&\approx {1\over 2}T(\bb)e_f^2 \alpha_{em}
\alpha_S(\bp_+^2) [z^2+(1-z)^2] G(\bDelta^2+Q_A^2)\nonumber\\
&&\times {Q_A^2 \over 
\pi (\bDelta^2 + Q_A^2 )^2} {\varepsilon^4 + (\bp_+^2)^2 \over (\varepsilon^2 + 
\bp_+^2)^4}
\, ,
\label{eq:10.20} 
\eea
where the integrated gluon structure function of the nucleon enters at
the factorization scale $\approx (\bDelta^2+Q_A^2)$. Neglecting the scaling 
violations in  $G(\bDelta^2+Q_A^2)$, we can evaluate  
expectation value $\langle \bDelta_{\perp}^2 \rangle$ subject to the 
condition $\bDelta^2 \lsim \bp_{+}^2$ as
\beq
\langle \bDelta_{\perp}^2 \rangle \approx  {Q_A^2 \over 2} 
\left( {Q_A^2 +\bp_+^2 \over \bp_{+}^2} \log{Q_A^2 +\bp_+^2 \over  Q_A^2} -1\right)
\label{eq:10.21}
\eeq
The differential out-of-plane momentum distribution can be evaluated 
as 
\beq
{d\sigma \over d\Delta_{\perp}} \approx 
{1\over 2} {Q_A^2 \over (Q_A^2 +\Delta_{\perp}^2)^{3/2}}
\label{eq:10.22}
\eeq
which shows that the probability to observe the away jet at $\Delta_{\perp}
\sim 0$ disappears $\propto 1/Q_A \propto 1/\sqrt{\nu_A}$.

The probabilistic character of our result (\ref{eq:10.17}) can be 
understood as follows. Hard jets
originate from small color dipoles. Their interaction with gluons
of the target nucleus is suppressed by screening of color charges
of the quark and antiquark in the color-singlet $q\bar{q}$ state which 
is manifest from the small cross section for a free nucleon target,
see eqs. (\ref{eq:2.9}), (\ref{eq:2.10}). The first inelastic 
interaction inside a nucleus converts the $q\bar{q}$ pair into
the color-octet state. The color charges of the quark and 
antiquark are no longer screened and rescatterings of the quark 
and antiquark in the color field of intranuclear nucleons are 
uncorrelated. Consequently, the broadening of the momentum 
distribution with nuclear thickness follows the probabilistic
picture.

% -------------------  Section 11

\section{Disappearance of jet-jet correlation in DIS off nuclear 
targets: minijets below the saturation scale}

In the above discusion of the single particle spectrum we 
discovered that the sea quarks evolve via  
the anticollinear, anti-DGLAP splitting of gluons 
into sea, when the transverse momentum of the parent gluons is larger
than the momentum of the sea quarks \cite{Saturation}, which 
suggests strongly
a complete azimuthal decorrelation of forward jets with the transverse
momenta below the saturation scale, $\bp_{\pm}^2 \lsim Q_A^2$. Our numerical
results reported in section 6 suggest that for the
realistic nuclei $Q_{A}^2$ does not exceed several GeV$^2$, hence 
this regime is a somewhat academic one. Still, let us assume that 
$Q_A$ is so large that jets with $\bp_{\pm}^2 \lsim Q_A^2$ are measurable.

Upon the slight generalization of (\ref{eq:10.16}) the distortion factor 
admits a representation
\bea
D_A(\bs,\br,\br',\bb)= 
\int_0^1 d \beta \exp\left\{-{1\over 2}[\beta \Sigma_1 + (1-\beta)\Sigma_2]
T(\bb)\right\} \, .
\label{eq:11.1}
\eea
It is still too complex if one is after the exact eigenvalues $\Sigma_{1,2}$,
however, it simplifies greatly if one invokes the large-$N_c$ approximation, 
in which case 
\beq
\Sigma_{2}= \langle 11| \sigma_4|11\rangle= \sigma(\br)
+\sigma(\br')
\eeq
  and
\bea
\Sigma_1 = \sigma(\bs)+\sigma(\bs+\br'-\br)\,.
\label{eq:11.2}
\eea
Because in the large-$N_c$ approximation $\Sigma_2=\langle 11| \sigma_4|11\rangle=
\sigma(\br)+\sigma(\br')$, the first two terms in the Sylvester
expansion (\ref{eq:3.19}) do vanish and 
only the last term will contribute to the
jet-jet inclusive cross section. In the large-$N_c$ approximation
there is only one transition from
the color-singlet to the color-octet state, once in the color octet state
the $q\bar{q}$ only oscillates in the color space. 
Then the different exponentials in 
the resulting distortion factor 
\bea
D_A(\bs,\br,\br',\bb)= 
\int_0^1 d\!\!\! \!\!\! \!\! \! &&\beta\,
\exp\left\{-{1\over 2}\beta[\sigma(\bs)+\sigma(\bs+\br'-\br)]
T(\bb)\right\} \nonumber\\
&& \times \exp\left\{-{1\over 2}(1-\beta)[\sigma(\br)+\sigma(\br')]
T(\bb)\right\} 
\label{eq:11.3}
\eea
admit a simple interpretation: The last two exponential factors describe
the intranuclear distortion of the incoming color-singlet $(q\bar{q})$
dipole state, whereas the former two factors describe the distortion of
the outgoing color-octet dipole states.

A straightforward generalization of (\ref{eq:10.16}) gives
\bea
D_A(\bs,\br,\br',\bb)= \int_0^1 d \!\!\!  \!\!\!  \!\!\!&&\beta \,
\int 
d^2\bkappa_1 \Phi((1-\beta)\nu_A(\bb),\bkappa_1)\exp(-i\bkappa_1 \br) 
\nonumber\\
&&\times \int d^2\bkappa_2 \Phi((1-\beta)\nu_A(\bb),\bkappa_2)\exp(i\bkappa_2 \br) 
\nonumber\\
&&\times \int d^2\bkappa_3 \Phi(\beta\nu_A(\bb),\bkappa_3)\exp[i\bkappa_3(\bs+\br'- \br)] 
\nonumber\\
&&\times \int d^2\bkappa_4 \Phi(\beta\nu_A(\bb),\bkappa_4)\exp(i\bkappa_4 \br) 
\label{eq:11.4}
\eea
and the jet-jet inclusive cross section takes the form
\bea
{d\sigma_{in} \over d^2\bb dz d\bp_{-} d\bDelta} &=&
{1\over 2(2\pi)^2} \alpha_S \sigma_0 T(\bb)\int_0^1 d \beta 
\int d^2\bkappa_1 d^2\bkappa_2 d^2\bkappa_3 d^2\bkappa 
f(\bkappa) \nonumber\\
&\times &\Phi((1-\beta)\nu_A(\bb),\bkappa_1) 
\Phi((1-\beta)\nu_A(\bb),\bkappa_2) \nonumber\\
&\times &
\Phi(\beta\nu_A(\bb),\bkappa_3) 
\Phi(\beta\nu_A(\bb),\bDelta -\bkappa_3 -\bkappa)\nonumber\\
&\times &\left\{\Psi(-\bp_{-} +\bkappa_2 +\bkappa_3)-
\Psi(-\bp_{-} +\bkappa_2 +\bkappa_3+\bkappa)\right\}^*\nonumber\\
&\times &
\left\{\Psi(-\bp_{-} +\bkappa_1 +\bkappa_3)-
\Psi(-\bp_{-} +\bkappa_1 +\bkappa_3+\bkappa)\right\}
\label{eq:11.5}
\eea
It is uniquely calculable in terms of the NSS-defined WW glue
of the nucleus.

Now consider the limiting case of minijets with the transverse
momentum below the saturation scale,
$|\bp_-|,|\bDelta| \lsim Q_A$. 
Notice, that $|\bkappa_{i}|\sim Q_A$, so that one can neglect
$\bp_-$ in the photon's wave functions and the decorrelation momentum
$\bDelta$ in the
argument of $\Phi(\beta\nu_A(\bb),\bDelta -\bkappa_3 -\bkappa)$.
The gross features of the product of the photon's wave functions
which enters the integrand of (\ref{eq:11.5}), after the averaging 
over $\bkappa_1$ and $\bkappa_2$ can be well approximated by 
\bea
&&\langle  \left\{ \Psi (-\bp_{-} +\bkappa_2 +\bkappa_3 )-
\Psi(-\bp_{-} +\bkappa_2 +\bkappa_3+\bkappa)\right\}^*
\nonumber\\
&&~ \left\{ \Psi (-\bp_{-} +\bkappa_1 +\bkappa_3)-
\Psi(-\bp_{-} +\bkappa_1 +\bkappa_3+\bkappa)\right\}\rangle_{\bkappa_{1,2}}
\nonumber\\ &&\Longrightarrow  | \Psi(\bkappa_3)-
\Psi(\bkappa_3+\bkappa)|^2
\label{eq:11.6}
\eea
The principal point is that the minijet-minijet inclusive cross 
section depends
on neither the minijet nor decorrelation momentum, which corroborates
the anticipated complete disappearance of the azimuthal decorrelation 
of jets with the transverse momentum below the saturation scale.

\section{Summary and conclusions}

We formulated the theory of DIS off nuclear targets based on the
consistent treatment of propagation of color dipoles in nuclear
medium. What is viewed as attenuation in the laboratory frame can
be interpreted as a fusion of partons from different nucleons of
the ultrarelativistic nucleus. Diffractive attenuation of color
single $q\bar{q}$ states gives a consistent definition of the 
WW unintegrated gluon structure function of the nucleus
\cite{NSSdijet,Saturation}. In these lectures we demonstrated how
all 
other nuclear DIS observables - sea quark structure function
and its decomposition into equally important genuine inelastic 
and diffractive components, exclusive diffraction off nuclei,
the jet-jet inclusive cross section,
- are uniquely calculable in terms of the NSS-defined 
nuclear WW glue. This
property can be considered as a new factorization which connects
DIS in the regimes of low and high density of partons. 

The two-plateau single parton (jet) inclusive cross section with the strong
$Q^2$ dependence of the plateau for truly inelastic DIS has not
been discussed before. A comparison with the initial state nuclear parton
densities which evolve from the NSS-defined WW nuclear glue shows
an exact equality of the FS and IS parton densities. The
plateau-like saturated nuclear quark density is suggestive of the
Fermi statistics, but our principal point that for any projectile
which interacts weakly with colored partons the saturated density
measures the momentum distribution in the $q\bar{q} , gg,...$ Fock
state of the projectile disproves the Fermi-statistics
interpretation. The spin and color multiplet of colored partons
the photon couples to is completely irrelevant, what only counts
is an opacity of heavy nuclei. 

The anti-collinear splitting of WW
nuclear glue into soft sea partons is a noteworthy feature of the
both diffractive DIS and IS sea parton distributions. The
collective nuclear WW gluon field exhibits strong nuclear suppression,
and the 
emergence of a saturated density of  sea partons from the
nuclear-diluted WW glue is due to the nuclear broadening of the
plateau. Because the predominance of diffraction is
a very special feature of DIS \cite{NZZdiffr}, one must be careful
with applying the IS parton densities to, for instance, nuclear
collisions, in which diffraction wouldn't be of any significance.
The anti-collinear splitting of WW nuclear glue is a clearcut
evidence for inapplicability of the DGLAP evolution to nuclear
structure functions unless $Q^2 \gg Q_{8A}^2$.  

Although in our derivation we focused on DIS, all the results for
hard single-jet and jet-jet inclusive cross sectinos are fully
applicable to production of jets in the beam fragmentation region
of  meson-nucleus collisions and can readily be extended
to nucleon-nucleus collisions. Indeed, as argued in \cite{NSSdijet},
the final state interaction between the final state quark and
antiquark can be neglected and plane-wave approximation becomes
applicable as soon as the invariant mass of the dijet exceeds a
typical mass scale of prominent meson resonances.

Of particular interest are the results on the azimuthal
decorrelation of hard jets, in particular, disappearance of
azimuthal jet-jet correlations of minijets with momenta below
the saturation scale. For the average DIS on heavy nuclei we
estimate $Q_{8A}^2 \approx 2$ GeV$^2$, but for the central 
DIS at a small impact parameter the saturation scale can be 
several times larger. For instance, from perihperal DIS to
central DIS on a heavy nucleus like $Pt$, $\nu_{8A}$ rises 
form $\nu_{8A}=1$ to $\nu_{8A}\sim 13$, so that according 
to (\ref{eq:10.22}) a probability to find the away jet
decreases by the large factor $\sim 3.5$ from peripheral
to central DIS off $Pt$ target.  

The early experience with application
of color dipole formalism to hard hadron-nucleus interactions
\cite{NPZcharm} suggests that our analysis can be readily
generalized to mid-rapidity jets. One only has to choose an
appropriate system of dipoles, for instance, the open heavy
flavor production can be treated in terms of the 
intranuclear propagation of the gluon-quark-antiquark system
in the overall color-singlet state. For this reason, we
expect similar strong decorrelation of mid-rapidity jets
in hadron-nucleus and nucleus-nucleus collisions. To this end,
recently the STAR collaboration reported a disappearance of
back-to-back high $p_{\perp}$ hadron correlation in central
gold-gold collisions at RHIC \cite{STARRHIC}. Nuclear
enhancement of the azimuthal decorrelation of the trigger 
and away jets discussed in sections 10 and 11 may 
contribute substantially to the STAR effect. Here we emphasize 
that we discuss the distortions of the produced jet-jet
inclusive spectrum due to interactions with the nucleons of the
target, a practical consideration of azimuthal decorrelations
in central heavy ion collisions must include rescatterings of
parent high-$p_{\perp}$ partons on the abundantly produced 
secondary hadrons.

This work has been partly supported by the INTAS grants 97-30494
\& 00-00366 and the DFG grant 436RUS17/89/02.\pagebreak\\

\end{document}